\documentclass[acmsmall, screen]{acmart}

\usepackage{enumitem}
\usepackage{multirow}
\usepackage{algorithm}
\usepackage{algorithmic}

\usepackage{subfigure}
\usepackage{makecell}

\AtBeginDocument{%
  \providecommand\BibTeX{{%
    \normalfont B\kern-0.5em{\scshape i\kern-0.25em b}\kern-0.8em\TeX}}}

\begin{document}


\title[MEGCF: Multimodal Entity Graph Collaborative Filtering for Personalized Recommendation]{MEGCF: Multimodal Entity Graph Collaborative Filtering for Personalized Recommendation}

\author{Kang Liu}
\email{kangliu1225@gmail.com}
\affiliation{
	\institution{Hefei University of Technology}
	\department{School of Computer Science and Information Engineering, Key Laboratory of Knowledge Engineering with Big Data, Intelligent Interconnected Systems Laboratory of Anhui Province}
	\streetaddress{485 Danxia Road}
	\city{Hefei}
	\state{Anhui Province}
	\country{China}
	\postcode{230601}
}

\author{Feng Xue}
\authornote{The corresponding author is Feng Xue}
\email{feng.xue@hfut.edu.cn}
\affiliation{%
	\institution{Hefei University of Technology}
	\department{School of Software, Key Laboratory of Knowledge Engineering with Big Data, Intelligent Interconnected Systems Laboratory of Anhui Province}
	\streetaddress{485 Danxia Road}
	\city{Hefei}
	\state{Anhui Province}
	\country{China}
	\postcode{230601}
}

\author{Dan Guo}
\email{guodan@hfut.edu.cn}
\affiliation{%
	\institution{Hefei University of Technology}
	\department{School of Computer Science and Information Engineering, Key Laboratory of Knowledge Engineering with Big Data, Intelligent Interconnected Systems Laboratory of Anhui Province}
	\streetaddress{485 Danxia Road}
	\city{Hefei}
	\state{Anhui Province}
	\country{China}
	\postcode{230601}
}

\author{Le Wu}
\email{lewu.ustc@gmail.com}
\affiliation{%
	\institution{Hefei University of Technology}
	\department{School of Computer Science and Information Engineering, Key Laboratory of Knowledge Engineering with Big Data, Intelligent Interconnected Systems Laboratory of Anhui Province}
	\streetaddress{485 Danxia Road}
	\city{Hefei}
	\state{Anhui Province}
	\country{China}
	\postcode{230601}
}

\author{Shujie Li}
\email{lisjhfut@hfut.edu.cn}
\affiliation{%
	\institution{Hefei University of Technology}
	\department{School of Software, Key Laboratory of Knowledge Engineering with Big Data, Intelligent Interconnected Systems Laboratory of Anhui Province}
	\streetaddress{485 Danxia Road}
	\city{Hefei}
	\state{Anhui Province}
	\country{China}
	\postcode{230601}
}

\author{Richang Hong}
\email{hongrc@hfut.edu.cn}
\affiliation{%
	\institution{Hefei University of Technology}
	\department{School of Computer Science and Information Engineering, Key Laboratory of Knowledge Engineering with Big Data, Intelligent Interconnected Systems Laboratory of Anhui Province}
	\streetaddress{485 Danxia Road}
	\city{Hefei}
	\state{Anhui Province}
	\country{China}
	\postcode{230601}
}

\renewcommand{\shortauthors}{Kang Liu, et al.}

\begin{abstract}
In most E-commerce platforms, whether the displayed items trigger the user's interest largely depends on their most eye-catching multimodal content.
Consequently, increasing efforts focus on modeling multimodal user preference, and the pressing paradigm is to incorporate complete multimodal deep features of the items into the recommendation module. 
However, the existing studies \textbf{ignore the mismatch problem between multimodal feature extraction (MFE) and user interest modeling (UIM)}. That is,
MFE and UIM have different emphases. 
Specifically, MFE is migrated from and adapted to upstream tasks such as image classification. In addition, it is mainly a content-oriented and non-personalized process, while UIM, with its greater focus on understanding user interaction, is essentially a user-oriented and personalized process.
Therefore, the direct incorporation of MFE into UIM for purely user-oriented tasks, tends to introduce a large number of preference-independent multimodal noise and contaminate the embedding representations in UIM.

This paper aims at solving the mismatch problem between MFE and UIM, so as to generate high-quality embedding representations and better model multimodal user preferences. Towards this end, we develop a novel model, \underline{m}ultimodal \underline{e}ntity \underline{g}raph \underline{c}ollaborative \underline{f}iltering, short for MEGCF. The UIM of the proposed model captures the semantic correlation between interactions and the features obtained from MFE, thus making a better match between MFE and UIM. More precisely, semantic-rich entities are first extracted from the multimodal data, since they are more relevant to user preferences than other multimodal information. These entities are then integrated into the user-item interaction graph. Afterwards,  a symmetric linear Graph Convolution Network (GCN) module is constructed to perform message propagation over the graph, in order to capture both high-order semantic correlation and collaborative filtering signals. Finally, the sentiment information from the review data are used to fine-grainedly weight neighbor aggregation in the GCN, as it reflects the overall quality of the items, and therefore it is an important modality information related to user preferences. Extensive experiments demonstrate the effectiveness and rationality of MEGCF\footnote{We release the complete codes of MEGCF at \textit{https://github.com/hfutmars/MEGCF}.}. 

\end{abstract}

\ccsdesc[500]{Information systems~Recommender systems}
\ccsdesc[500]{Information systems~Personalization}
\keywords{Collaborative Filtering, Semantic Correlation, Multimodal User Preference, Multimodal Semantic Entity, Collaborative Signal, Graph Convolutional Network, Sentiment Analysis}

\maketitle

\section{INTRODUCTION}\label{sec:introduction}
The personalized recommender algorithm plays a crucial role in many online services, such as E-commerce, content-sharing platform, and social media. 
Collaborative Filtering (CF)\cite{scf} is the most widely used recommender method, which assumes that there is a correlation signal between observed user-item pairs through collaborative relationships, and the signal enables accurate assessment of users' preference over items, which is referred to as CF signal in some studies\cite{kgat,ngcf}.
However, CF faces the challenges of sparsity and cold-start, that is, 
the inability to capture sufficient CF signals from sparse interactions to generate high-quality recommendations.  Content-enriched recommender methods can efficiently alleviate this problem by drawing on additional information of users and items, such as demographic features\cite{svdfeature}, attributes of items\cite{DeepFM-ijcai2017}, social relationships\cite{DiffNet-sigir2019}, knowledge graphs \cite{kgcn}, and multimodal content (\textit{e.g.}, images, short videos, titles, reviews, \textit{etc.}) of items \cite{mmgcn} (referred to as multimodal recommender method), in order to enrich the representations of users and items.
This work focuses on the research of multimodal recommender methods due to the following two considerations: (1) in most recommendation scenarios, multimodal information is the dominant presentation of the item and it  directly engages with users. Therefore it contains abundant user preference-related clues that differ from the collaborative relationships in the interactions; and
(2) the recent success of video-sharing platforms, such as Tiktok and Kwai, bring increasing attention to extracting user preference over multimodal content.

The existing multimodal recommender efforts can be broadly categorized into two types of frameworks: {Separated Framework (SF)} and {End2end Framework (EF)}. They are both equipped with two main modules: \textbf{multimodal feature extraction (MFE)} and \textbf{user interest modeling (UIM)}. In the SF-based methods\cite{vbpr,CKE-kdd2016,mmgcn,mgat-ipm2020}, the MFE module uses a pre-trained network migrated from the upstream task to extract the full range of multimodal deep features. In addition, the UIM module incorporates these deep features into the user preference modeling. Compared with SF, EF-based methods\cite{comparativedl,DVBPR-icdm2017,ConvMF-recsys2016,MRG-www2019,DeepCoNN-wsdm2017} differ by fusing the MFE and UIM modules into an end2end framework and using interaction data to jointly optimize them. 
Note that a more comprehensive overview of these multimodal recommender methods is provided in Section \ref{subsec:mmrec}.

\begin{figure}
	\centering
	\includegraphics[height=7.52cm, width=12cm]{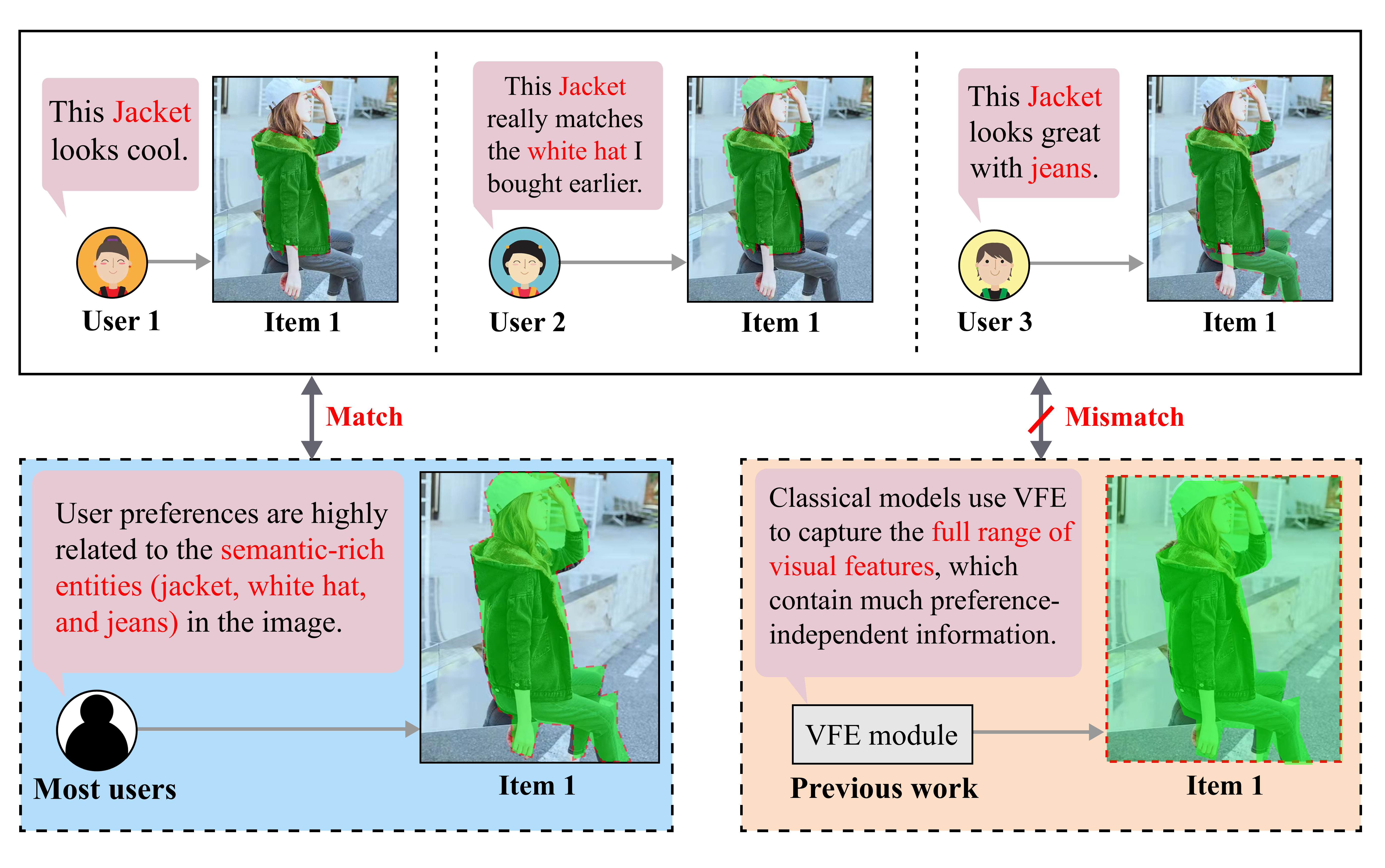}
	\caption{An example to illustrate the mismatch between user preferences and the visual feature extraction (VFE) module when processing visual content, where \textit{VFE} is generally a convolutional neural network pre-trained on a large-scale visual dataset; \textit{User 1}, \textit{User 2}, and \textit{User 3} are users who interact with \textit{Item 1} together.}
	\label{fig:motivation-1}
\end{figure}

Although the existing studies demonstrate the effectiveness of this schema, they ignore the mismatch problem between the MFE and UIM modules.
Specifically, MFE, as a module tailored for upstream tasks, aims at mining multimodal deep features that are relevant to the specific upstream task but not to the user preferences. Therefore, it is a content-oriented process. On the contrary, UIM, as the core of recommendation task, is a user-oriented and personalized module which aims at collecting and then processing preference-related features. 
In general, MFE and UIM have completely different emphases. In other words, they are mismatched in user preference inference, thus limiting the positive impact of multimodal information.
In addition, the features output by the MFE module contain a large amount of preference-independent noisy data, resulting in contamination of the embedding representations. Figure \ref{fig:motivation-1} shows a specific mismatch phenomenon between user preferences and visual feature extraction (VFE) module. That is, whether or not a user will purchase an item is highly related to the semantic-rich entities ($e.g.$, jacket, white hat, and jeans) in the image of this item, while the VFE module captures the full range of visual features that contain a large number of preference-independent information ($e.g.$, background, brightness, and the relative position of the entities).

In order to solve the mismatch problem, it is fundamental to transform content-oriented MFE into user-oriented one, that is, mining preference-related information and filtering out preference-independent noisy data. In practice, semantic-rich entities in multimodal content are highly correlated to the user purchase behavior ($cf$. Figure \ref{fig:motivation-1}). Therefore, extracting these semantic entities rather than the full multimodal deep features facilitates the transformation of MFE. Furthermore, the user sentiment information contained in item reviews is a crucial and typical preference-related feature of textual modality, as it reflects the overall quality of the item, and the users always tend to purchase high-quality items. Consequently, capturing sentiment information can further transform the MFE module into user-oriented one.

\begin{figure}
	\centering
	\includegraphics[height=4.134cm, width=8cm]{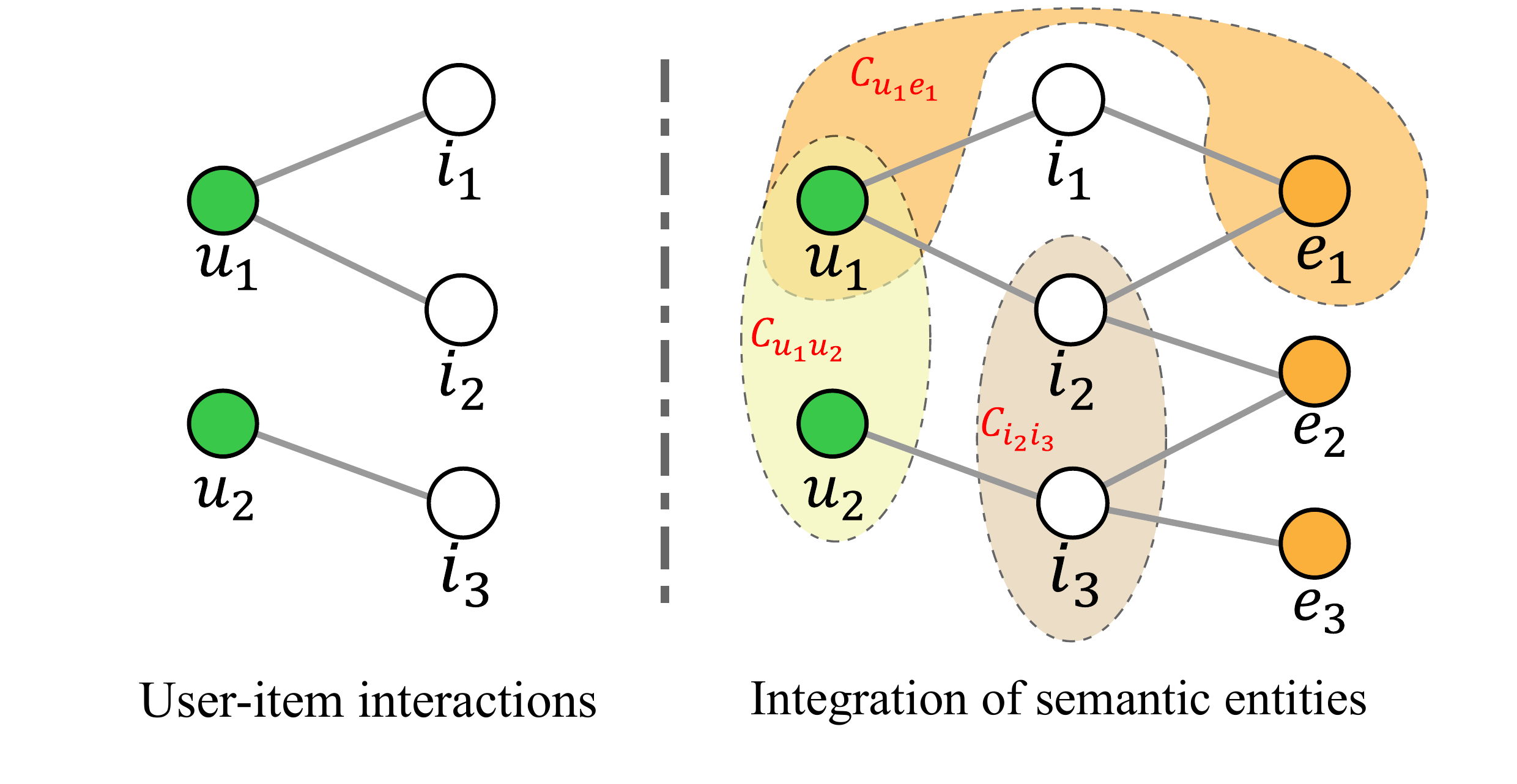}
	\caption{Illustration of multimodal semantic correlation, where $u$, $i$, and $e$ denote the user, item, and semantic entity, respectively, and $C_{ab}$ denotes the semantic correlation between $a$ and $b$.}
	\label{fig:motivation-2}
\end{figure}

After achieving the transformation of MFE, the next step is to establish an association between MFE and UIM. Methodologically, EF is a reasonable option as it uses interaction data to jointly optimize MFE and UIM. However, in fact, multimedia recommendations are generally applied in sparse and cold-start scenarios, which means that the large number of learnable parameters in the MFE module are difficult to be optimized. From this view, SF is considered as the overall framework, and the MFE is treated as a pre-processing module. Moreover, multimodal semantic correlation, which seamlessly associates the MFE and UIM through interaction relationship, is proposed. Figure \ref{fig:motivation-2} presents a simple example to show the multimodal semantic correlation and its importance for modeling user preferences. The left subfigure shows that the interaction sparsity makes the measurement of the similarity between nodes difficult (as there is no path between $u_1$ and $u_2$). In the right subfigure, after incorporating the semantic entities, the similarity between $u_1$ and $u_2$ can be accurately measured due to the fact that a path <$u_1$,$i_2$,$e_2$,$i_3$,$u_2$> emerges between them (the similarity is referred to as the multimodal semantic correlation $C_{u_1u_2}$ between $u_1$ and $u_2$). Similarly, $C_{i_1i_2}$ represents the semantic correlation between $i_1$ and $i_2$. In addition, semantic correlation can also represent the user preference for entities ($e.g.$, $C_{u_1e_1}$). $e_1$ is clearly the best match for $u_1$'s preference, as the most paths between them exist.

In order to better quantify the aforementioned semantic correlation, the Graph Convolution Network (GCN)\cite{gcn} is the optimal choice as the recent study \cite{ngcf, light} has demonstrated its outperformance  in capturing high-order correlation between nodes on graph. In addition, the modeling of this correlation can be further enhanced by fine-grained weighting of neighbor aggregation in GCN, which is conventionally implemented by constructing attention networks \cite{gat}. However, in the multimodal recommendation scenarios, the weighting strategy of self-attention is suboptimal, because it ignores the capture of preference-related sentiment information (or overall quality) hidden in the items (evidence in Section \ref{subsubsec:ex:sentiment-gat}). From this view, it is a better option to mine sentiment information from item reviews and then use it to weight neighbor aggregation. Moreover, the weights obtained by sentiment information are static parameters, and therefore they do not increase the training difficulty and computational burden of the model.

Generally speaking, in this paper, we propose a GCN-based multimodal recommender method, referred to as MEGCF. First, we employ advanced deep learning techniques to mine semantic entities from multimodal content and seamlessly integrate them into the user-item interaction graph. Next, we construct a symmetric linear GCN module to perform high-order message propagation on the graph, thus modeling multimodal semantic correlation among nodes and extends it to higher-order. Finally, in order to make full use of preference-related features in textual modality, we utilize sentiment analysis techniques to extract sentiment features from item reviews and proposed a sentiment weighting strategy to enhance the graph convolution operations.
We conduct extensive experiments on three public datasets, and the results show that our proposed method significantly outperforms the state-of-the-art multimodal recommender method (GRCN\cite{grcn}) and GCN-based CF method (SGL\cite{sgl-sigir2021}). Furthermore, we validate the effectiveness of each component in MEGCF through sufficient ablation experiments. 

we summarize the contributions of this paper as follows:
\begin{itemize}[leftmargin=*]
\item We highlight the mismatch problem between multimodal feature extraction and user interest modeling in existing multimodal recommender methods, which contaminates embeddings and makes the models non-robust. To solve this problem, we propose to model multimodal semantic correlation and extract sentiment information.
\item We propose a novel GCN-based multimodal recommender method, referred to as MEGCF, which utilizes multimodal semantic entity extraction and sentiment-weighted symmetric linear GCN module to achieve simultaneous capture of high-order multimodal semantic correlation and CF signals.
\item We conduct extensive experiments on three real-world datasets to validate the state-of-the-art performance of MEGCF. In addition, further ablation experiments are performed to verify the effectiveness of each component in MEGCF. To facilitate subsequent research, we release the complete code and data of MEGCF at \textit{https://github.com/hfutmars/MEGCF}.
\end{itemize}

\section{RELATED WORK}
 {In this section, we briefly review three types of recommender methods that are most relevant to our work: traditional Collaborative Filtering (CF)-based methods, Graph Convolution Network (GCN)-based recommender methods, and multimodal recommender methods.}
\subsection{Traditional CF-based Recommender Methods}
Since the base framework of our work is the CF approach, we present CF and related work based on it here.
CF\cite{scf} assumes that similar users exhibit similar interests in items they historically interacted with. Matrix Factorization (MF)\cite{mf} is the pioneering work of CF, which generates dense embedding representations for users and items by mapping their ID information, and reconstructs the unobserved interactions between users and items with the inner product of their embeddings. BiasSVD \cite{biassvd} assumes that MF fails to accurately capture the difference in preferences between users (or items) and introduces bias terms on top of MF to compensate for the weakness of embedding expressiveness. 
However, the model performance of MF and BiasSVD strongly depends on sufficient interaction data, that is, they suffer from the problem of sparsity and cold-start.
To tackle this problem, one promising solution is to leverage side information that may preserve clues related to user preferences. 
SVD++\cite{svd++} and FISM\cite{fism} incorporate collections of items that users have historically interacted with into user embeddings, and their effectiveness in explicitly modeling interactions have been validated in subsequent work\cite{ngcf}\cite{light}.
Unlike merely using ID information, another family of CF methods focuses on mining user preferences from interaction-independent information. For example, SVDFeature\cite{svdfeature} is a feature-based MF model that incorporates user and item attributes into the embeddings to enhance their expressiveness. The knowledge graph-based recommender methods \cite{kgcn}\cite{ripplenet} extract item-related attribute entities and relations from external knowledge graphs into the item embeddings, thus achieving the association between user preferences and knowledge.
In contrast to the above CF methods centered on enhancing embedding, NCF\cite{ncf} and DICF\cite{dicf} use deep neural networks instead of simple inner products, enhancing the modeling of complex interactions.

\subsection{GCN-based Recommender Methods}
In Section \ref{subsec:slgcn}, we propose an improved GCN module for extending low-order features to higher-order. Thus we present GCN-based recommender methods here.
GCN\cite{gcn}\cite{graphsage} is a deep neural network proposed for graph-structured data. In recent years, it has been widely studied and led to satisfactory results in recommender systems.
The core paradigm of GCN is to iteratively aggregate neighbor nodes into the embeddings of the target nodes, thus explicitly capturing important high-order connectivities on the graph. 

To the best of our knowledge, GC-MC\cite{gcmc} is the first approach that applies GCN to recommender systems, which leverages graph convolution to aggregate one-hop neighbor nodes into the embeddings of the target nodes. 
PinSAGE\cite{pinsage} and NGCF\cite{ngcf} integrate high-order neighbor nodes into the embedding generation of the target node. Specifically, PinSAGE combines random walks and GCN to achieve embedding generation on large-scale item-item graphs, demonstrating that GCN-based methods can be efficiently applied to web-scale recommendation scenarios.
NGCF\cite{ngcf} is a recommender framework which combines GCN and MF. It uses a layer aggregation mechanism\cite{layer-aggregation} to concatenate the embeddings of all the layers as the final node representations, in order to capture the semantic information preserved by different graph convolution layers.
Some subsequent studies\cite{rgcf,lrgccf,light} deduced that the simplified GCN\cite{sgc} is better suited for modeling interactions in recommendation scenarios.
LR-GCCF\cite{lrgccf} and LightGCN\cite{light} can be broadly seen as the lightweight versions of NGCF. LRGCCF removes the nonlinear activation function from NGCF and re-analyzes the layer aggregation mechanism in NGCF from a residual perspective. LightGCN removes both the nonlinear activation function and the weight transformation matrices from NGCF. In addition, extensive experiments show that a more concise structure of LightGCN can achieve a better performance. For convenience, we refer to the GCN module in LightGCN as the linear GCN in this work. 
Essentially, the Graph Laplacian Norm in GCN can be considered as a scaling of node degree (or popularity) at a fixed granularity, while users have different sensitivities to popularity features. Consequently, JMPGCF\cite{jmpgcf} is proposed to construct different Graph Laplacian Norms for GCNs, in order to capture multi-grained popularity features, and thus better model user preferences on popularity.
Although the previously mentioned GCN-based approaches achieve a high performance by explicitly modeling high-order connectivities, this schema of graph convolution ignores the modeling of the diversity of user intents.
Therefore, DGCF\cite{DGCF-sigir2020} is proposed to construct a graph disentangling module, in order to iteratively refine the intent-aware interaction graphs and factorial representations.
Some researchers have recently tried to fuse contrastive learning and GCN to implement recommendation. For instance, SGL\cite{sgl-sigir2021} is a model-agnostic self-supervised contrastive learning framework which incorporates node self-discrimination task into recommendation module and jointly learn them. Thus, it alleviates the long-tail problem and enhances the model robustness to noisy interactions.

{It is important to mention that the GCN module in the proposed MEGCF is significantly different from the above-mentioned GCN in terms of graph convolution operations and overall structure. More precisely, we incorporate sentiment weighting strategy and popularity features based on linear graph convolution. In addition, for the overall structure, we use two symmetric versions of this GCN to construct the final module for embedding generation.}

\subsection{Multimodal Recommender Methods}\label{subsec:mmrec}
{Considering that the motivation of this study is to address the mismatch problem between multimodal feature processing and user preference modeling, $i.e.$, it focuses on how to improve the utilization of multimodal data in recommendation scenarios. Thus,  the existing work on multimodal recommendations\cite{mm-rs-recsys2021} is presented.
Multimodal recommender methods can be considered as content-enriched ones \cite{content-rs}, that leverage multimodal content to assist in the recommendation. 
The overall framework of the existing multimodal recommender efforts can be roughly divided into two categories: \textbf{Separated Framework (SF)} and \textbf{End2end Framework (EF)}. SF separates the two modules, multimodal feature processing and user preference modeling, and uses the multimodal features obtained in the former module to enrich the embedding representation in the latter module. On the contrary, EF fuses these two modules and jointly learns the two tasks of multimodal feature processing and user preference modeling in order to perform complementarity between them.}

{First, we introduce the SF-based multimodal recommender methods.} VBPR\cite{vbpr} is an early approach which uses Convolution Neural Network (CNN) pre-trained on ImageNet to extract deep visual features of images and incorporate them into feature representations of items in the MF framework.
VPOI\cite{VPOI-www2017} incorporates pre-extracted deep visual features into the PMF\cite{PMF-nips2007} framework in order to achieve POI check-in recommendations.
Inspired by the fact that user preferences tend to exhibit significant variability across different modalities,
increasing efforts focus on simultaneously capturing clues of user preferences on multiple modalities.
For instance, CKE\cite{CKE-kdd2016} incorporates knowledge graphs, visual features, and textual features into embedding representation. MMGCN\cite{mmgcn} is a GCN-based multimodal recommender method, which constructs three GCN modules to model user preference on visual, textual, and audio modalities, respectively. Compared with CKE, MMGCN can mine higher-order multimodal similarity features, thus achieving better recommendation results.
Based on MMGCN, MGAT\cite{mgat-ipm2020} constructs the attention network to adaptively calculate the weights of user preferences on different modalities.
MKGAT\cite{mkgat-cikm2020} leverages the multimodal contents to construct a multimodal knowledge graph. In addition, the tail nodes of the knowledge graph are dense vectors represented by fusing the deep features of different modalities.
These tail nodes are referred to as multimodal entities in MKGAT, which is fundamentally different from the proposed multimodal semantic entities that do not use deep features but extract semantic-rich entities from multimodal content.
A recent GCN-based multimodal recommender method, HUIGN\cite{huign}, which constructs a hierarchical graph structure and designs two types of information aggregation modules ($i.e.$, intra-level and inter-level aggregation) to model multi-level user intents. Therefore, it ensures the generation of high-quality user and item representations.

{Afterwards, we introduce EF-based multimodal recommender methods.}
DVBPR\cite{DVBPR-icdm2017} is an extended version of VBPR which integrates the CNN module into the MF module to jointly train image representations and recommendation modules in an end-to-end manner.
ConvMF\cite{ConvMF-recsys2016} is a fusion model of CNN and MF, in which the CNN module captures the contextual information of item reviews. Therefore, it improves the prediction accuracy. 
DeepCoNN\cite{DeepCoNN-wsdm2017} constructs two parallel CNN modules to learn the behavioral features of users and the attribute features of items from user-related and item-related reviews, respectively.
MRG\cite{MRG-www2019} is a multi-task learning model which models both the review generation module and the rating prediction module. It also jointly learns both modules to better mine user preferences on textual modality.
Methodologically, the EF-based approach incorporates interaction data into the multimodal feature processing to guide the mining of  preference-related multimodal features, which is supposed to be stronger than the SF-based approach. However, we argue that they still suffer from some limitations. Specifically, multimodal recommender methods are mostly applied in sparse and cold-start recommendation scenarios, which means that the large number of learnable parameters introduced by EF are difficult to be optimized.
In addition, integrating multimodal feature processing modules on general recommendation modules degrades the efficiency of model training and inference. Therefore, in this paper, we adopt the idea of SF to implement MEGCF.

{Despite the progress of these works, they all essentially use complete multimodal deep features to participate in the feature representations of items while ignoring the mismatch problem between multimodal feature processing and user preference modeling, which results in embedding contamination.}

\section{METHODOLOGY}\label{sec:method}
In this section, we present the overall framework of the MEGCF model shown in Figure \ref{fig:model}, which can be divided into three main components: (1) the multimodal semantic entity extraction layer for extracting semantic-rich entities from multimodal data, and then incorporating them into the user-item interaction graph; (2) the sentiment-weighted symmetric linear Graph Convolution Network (GCN) module for capturing both high-order multimodal semantic correlation and CF signal during embedding generation; and (3) the model prediction\&optimization layer for estimating the preference scores of user-item pairs and updating model parameters. 

\subsection{Multimodal Semantic Entity Extraction Layer}\label{subsec:mmem}
In order to reduce the negative impact of redundant information in multimodal data with low relevance to user preferences, we propose to extract semantic entities from multimodal data and associate them with items, as semantic entities have the potential to be more interest-provoking for users than other modal information such as background, position, angle, brightness, \textit{etc.} In the following, we detail the specific extraction of semantic entities for visual and textual modalities, respectively, and utilize them and the user-item interaction graph to construct a collaborative multimodal interaction graph.

\subsubsection{\textbf{Visual Semantic Entities Extraction}}
In order to ensure efficiency in entity extraction, we use the technique of image classification rather than object detection to extract visual semantic entities from images. Specifically, for an item $i$, we feed its corresponding image into a PNASNet model \cite{PNASNet-eccv2018} (an advanced image classification method) pre-trained on the ImageNet dataset. The model then outputs a probability distribution over 1000 categories, and we take the top-ranked categories
as the semantic entities in the image. These entities can also be understood as the most likely sub-objects present in the image. Finally, we perform the above operations on all the items to obtain $\mathcal{E}_V$, which is the set of semantic entities on the visual modality.

\subsubsection{\textbf{Textual Semantic Entities Extraction}}
The textual data of an item $i$ consist of a title and some reviews, on which we first perform pre-processing operations, including special characters removal, words segmentation, and stop words removal. Afterwards, as the title is informative and objective, we directly use the pre-processed words as textual entities in the title. Reviews express users' experiences and sentiment about the item $i$, and contain relatively less information. In addition, they are more subjective than titles. Therefore, we use the SGRank model \cite{SGRank-acl2015} (an advanced keyword extraction method) to further extract keywords from the pre-processed reviews, and its outputs are the semantic entities in the reviews. Finally, we iteratively perform the above operations on all the items in order to obtain the set of semantic entities on the text modality, $\mathcal{E}_T$.

Compared with the direct utilization of full multimodal deep features, semantic entity extraction can better reduce the preference-independent multimodal content. However, this approach still has limitations in entity detection accuracy. More precisely, when the neural network model pre-trained in the upstream task is directly migrated to the recommendation task, there is a significant degradation in entity detection accuracy (wrong or missing detection) occurs. Empirically, fine-tuning the pre-trained model using partially labeled data from the recommendation task is methodologically feasible. However, the multimodal data in the recommendation dataset are unlabeled. Despite the problem of wrong or missing detection, the proposed MEGCF is significantly stronger than most of the multimedia recommender methods ($cf.$ Table \ref{tab:overall}). In addition, there is a high probability that MEGCF will be further enhanced if we can improve the accuracy of entity extraction. We leave it for future considerations.

\begin{figure}
	\centering
	\includegraphics[height=8.43cm, width=14cm]{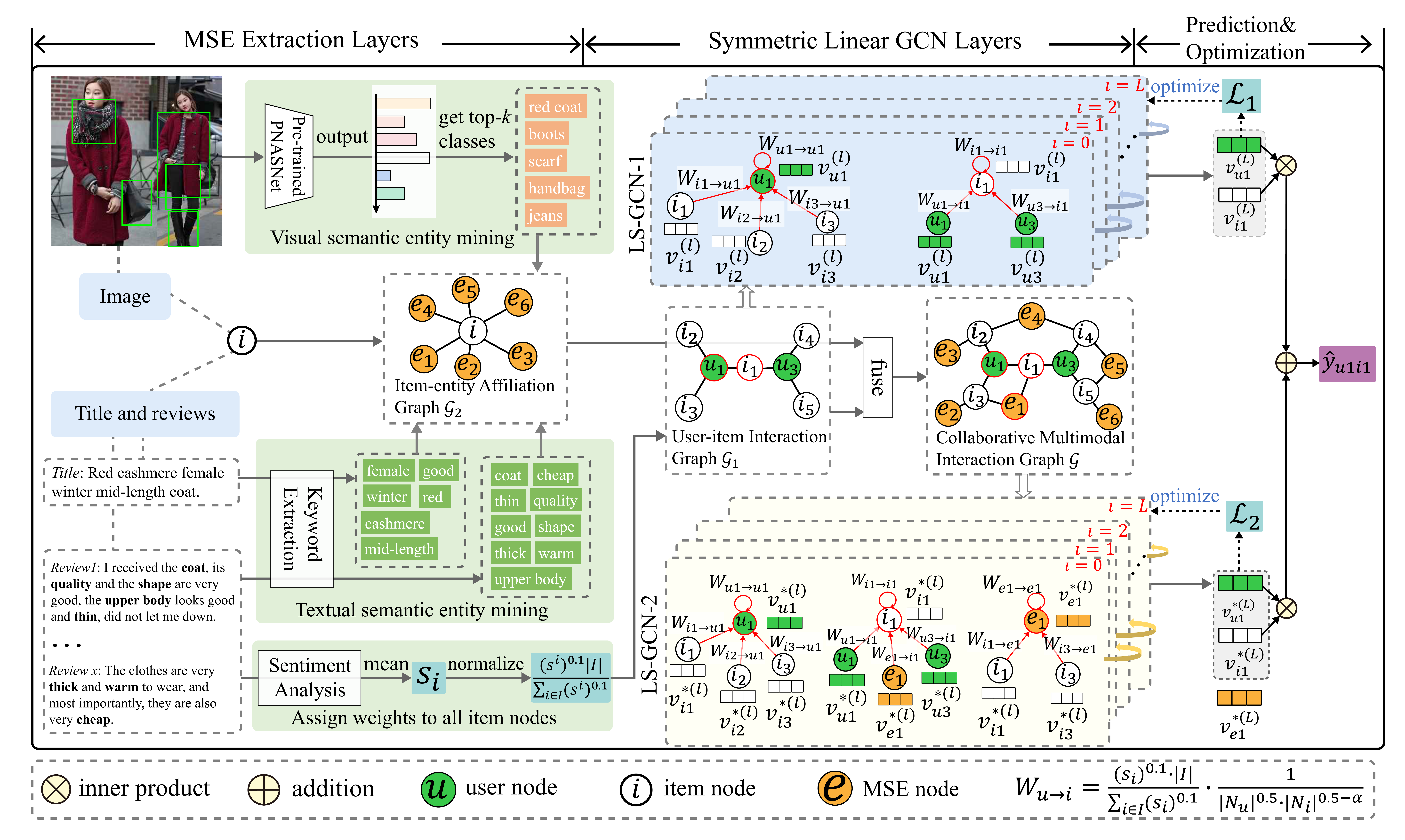}
	\caption{Illustration of the proposed MEGCF. The target user and item are $u_1$ and $i_1$, MSE denotes multimodal semantic entity, and $L$ is the max number of graph convolution layers. }
	\label{fig:model}
\end{figure}

\subsubsection{\textbf{Collaborative Multimodal Interaction Graph}}\label{subsubsec:cmig}
First, user-item interactions can be translated into a bipartite graph structure $\mathcal{G}_1=\{(u, r_{ui},i)|u \in \mathcal{U}, i\in \mathcal{I}\}$, where $\mathcal{U}$ and $\mathcal{I}$ are the user and item sets, respectively, $r_{ui}=1$ denotes that there is an interaction between user $u$ and item $i$, otherwise $r_{ui}=0$. Then, based on the affiliation between item $\mathcal{I}$ and the multimodal entities $\mathcal{E}=\mathcal{E}_V \cup \mathcal{E}_T $, we construct an item-entity bipartite graph $\mathcal{G}_2=\{(i, r_{ie}, e)|i\in\mathcal{I},e\in\mathcal{E}\}$, where $r_{ie}=1$ denotes that the entity $e$ is extracted from the multimodal data of item $i$, otherwise $r_{ie}=0$. Finally, as shown in Figure \ref{fig:model}, the item nodes are used as a bridge to fuse these two bipartite graphs into a new user-item-entity tripartite graph, $\mathcal{G}=\{(u,r_{ui},i),(i,r_{ie},e)|u\in\mathcal{U},i\in\mathcal{I},e\in\mathcal{E}\}$, named collaborative multimodal interaction graph. 

\subsection{Sentiment-weighted Symmetric Liner GCN Layer}\label{subsec:slgcn}
We design a sentiment-weighted symmetric linear graph convolution block to perform message propagation on the user-item graph $\mathcal{G}_1$ and the collaborative multimodal interaction graph $\mathcal{G}$, in order to capture the high-order CF signal and multimodal semantic correlation, respectively. In the following, we detail the process of embedding generation in this block.

\subsubsection{\textbf{Embedding Initialization}}\label{subsubsec:ei}
Following the ID embedding-based recommender models, we initialize all the users, items, and semantic entities by mapping their IDs to the corresponding dense low-dimensional vector representations as follows:
\begin{equation}\label{eq:initial}
\mathcal{V} = \{v_{u_1}^{(0)},...,v_{u_{|\mathcal{U}|}}^{(0)}, v_{i_1}^{(0)},...,v_{i_{|\mathcal{I}|}}^{(0)}, v_{e_1}^{(0)},...,v_{e_{|\mathcal{E}|}}^{(0)}\}, 
\end{equation}
where $\mathcal{V} \in \mathbb{R}^{{(|\mathcal{U}| + |\mathcal{I}| + |\mathcal{E}|)} \times d}$ is the embedding matrix of all the nodes in $\mathcal{G}$, $|\mathcal{U}|$, $|\mathcal{I}|$, and $|\mathcal{E}|$ are the number of users, items, and semantic entities, respectively, $d$ is the dimension length of embeddings. It is important to mention that the user and item nodes in $\mathcal{G}_1$, $\mathcal{G}_2$ and $\mathcal{G}$ are parameter shared.

\subsubsection{\textbf{Review-based Sentiment Extraction}}
As the reviews are highly subjective and can reflect the users' sentiments towards the target item, we propose to use an advanced sentiment analysis technique \cite{Senta-acl2020} to extract sentiment information from reviews, which allows fine-grained weighting to items. We formulate the calculation of sentiment score for item $i$ as follows:
\begin{equation}\label{eq:senti}
	s_i = \frac{\sum_{t \in T_i}{f(t)}}{|T_i|},
\end{equation}
where $T_i$ is the review set for item $i$, $|T_i|$ denotes the size of $T_i$, $f(\cdot)$ represents the pre-trained SENTA model\cite{Senta-acl2020}, which outputs the sentiment score from the input review, and $s_i$ is the mean sentiment score for item $i$.

In subsequent message propagation, we use $s_i$ to weight all the item nodes in $\mathcal{G}_1$ and $\mathcal{G}$.

\subsubsection{\textbf{Sentiment-weighted Embedding Generation}}\label{subsubsec:embed-generation}

The average sentiment in an item's review directly reflects the overall quality of the item. Most users usually buy high-quality items, $i.e.$, there is a higher affinity between user node and high-quality item node. Consequently, we propose to use the average sentiment score to distinguish the importance of different neighbors during the graph convolution process, so that the overall quality information mined from item reviews can be incorporated into the node representation.  
Note that the Graph Laplacian Norm in GCNs essentially uses node degree (number of interactions or popularity) to assign weights for neighbor aggregations, which is significantly different from the review-based sentiment weights. Therefore, using both the sentiment weights and Graph Laplacian Norm is a better option.
It is worth mentioning that GAT\cite{gat}, which is a variant of GCN, uses a self-attention mechanism to weight neighbor aggregations. We argue that the weighting strategy of GAT is suboptimal under the multimodal recommendation scenarios, because it only uses interactions and ignores the capture of sentiment information (or overall quality) inherent in the items (evidence in Section \ref{subsubsec:ex:sentiment-gat}).
In addition, Linear GCNs can better capture high-order CF signals, compared with the traditional nonlinear GCNs (evidence in \cite{light,rgcf}).


Based on these considerations, we devise a linear sentiment-weighted GCN structure to perform message propagation on the user-item interaction graph $\mathcal{G}_1$, in order to better capture high-order CF signals. For convenience, we refer to this structure as \textbf{LS-GCN-1}. For the target user $u1$ and item $i1$, we formulate their embedding generation in LS-GCN-1 as follows:
\begin{equation}\label{eq:i_embed_1}
	v_{i1}^{(l)} =  \sum_{u \in N_{i1}\cup i1}{\frac{(s_{i1})^{\gamma}|\mathcal{I}|}{\sum_{i \in \mathcal{I}}{(s_i)^{\gamma}}} \cdot \frac{1}{|N_{i1}|^{0.5}|N_u|^{0.5-\alpha}} \cdot v_u^{(l-1)}},
\end{equation}

\begin{equation}\label{eq:u_embed_1}
v_{u1}^{(l)} =  \sum_{i \in N_{u1}\cup u1}{\frac{(s_{i})^{\gamma}|\mathcal{I}|}{\sum_{i \in \mathcal{I}}{(s_i)^{\gamma}}} \cdot \frac{1}{\left|N_{u1}\right|^{0.5}\left|N_i\right|^{0.5-\alpha}} \cdot v_i^{(l-1)}},
\end{equation}
where $l$ denotes the number of current graph convolution layers, $N_u$ and $N_i$ are the neighbor nodes of user node $u$ and item node $i$ in $\mathcal{G}_1$, respectively, $|N_u|$ and $|N_i|$ denotes the size of $N_u$ and $N_i$, respectively, $\frac{(s_{i})^{\gamma}|\mathcal{I}|}{\sum_{i \in \mathcal{I}}{(s_i)^{\gamma}}}$ is the weight allocated to item $i$ using the sentiment score $s_i$ in Equation \ref{eq:senti}, which can also be considered as a normalization on the sentiment score $s_i$, $|\mathcal{I}|$ denotes the number of items, and $\gamma$ is used to smooth the sentiment score $s_i$ (we set $\gamma = 0.1$ in experiments for simplicity), note that we set $s_i=1.0$ when $i=u1$ in Equation \ref{eq:u_embed_1}.
In addition, following the recent graph learning based method\cite{jmpgcf}, we fine-tune the classical Graph Laplacian Norm $\frac{1}{|N_{i}|^{0.5}|N_u|^{0.5}}$ to the form of $\frac{1}{|N_{i}|^{0.5}|N_u|^{0.5-\alpha}}$, where $\alpha \in (0, 0.5)$ is a hyper-parameter used to control the model sensitivity to popularity information. For convenience, we term this improved Graph Laplacian Norm as popularity-aware norm (short for PN, Section \ref{subsubsec:g} validates the effectiveness of PN). Specifically, when $\alpha$ is larger, the embedding value obtained from the graph convolution is larger, which corresponds to the fact that the model is more sensitive to the popularity.

By multimodal semantic entity extraction and the construction of multimodal collaborative interaction graphs, we can model important semantic correlation, and thus reduce the negative impact of preference-independent multimodal information.
Now we move forward to capture this correlation and extend it to higher-order. We propose to construct a linear sentiment-weighted GCN structure similar to LS-GCN-1, referred to as \textbf{LS-GCN-2}. More precisely, we use LS-GCN-2 to iteratively perform message propagation over the collaborative multimodal interaction graph $\mathcal{G}$, in order to bridge multimodal entity information to user representations using item nodes, while user-item interactions can, in turn, facilitate the learning of multimodal semantic correlation.

We first formulate the embedding output at $l$-th layer in LS-GCN-2 for the target user node $u1$ as follows:
\begin{equation}\label{eq:u_embed_2}
{v^*}_{u1}^{(l)} =  \sum_{i \in N_{u1}\cup u1}{\frac{(s_{i})^{\gamma}|\mathcal{I}|}{\sum_{i \in \mathcal{I}}{(s_i)^{\gamma}}} \cdot \frac{1}{\left|N_{u1}\right|^{0.5}\left|N_i\right|^{0.5-\alpha}} \cdot {v^*}_i^{(l-1)}},
\end{equation}
where ${v^*}_{i}^{(0)}$ is the initialized embedding of $i$ in Equation \ref{eq:initial}, which is equivalent to ${v}_{i}^{(0)}$, note that ${v^*}_{i}^{(l)}$ is inherently different from ${v}_{i}^{(l)}$ in Equation \ref{eq:u_embed_1} because ${v^*}_{i}^{(l)}$ is incorporated with the message from multimodal semantic entities.

We then present the embedding generation for the target entity $e1$ as follows:
\begin{equation}\label{eq:e_embed_2}
{v^*}_{e1}^{(l)} =  \sum_{i \in N_{e1}\cup e1}{\frac{(s_{i})^{\gamma}\left|\mathcal{I}\right|}{\sum_{i \in \mathcal{I}}{(s_i)^{\gamma}}} \cdot \frac{1}{\left|N_{e1}\right|^{0.5}\left|N_i\right|^{0.5-\alpha}} \cdot {v^*}_i^{(l-1)}},
\end{equation}
note that we set $s_i=1.0$ when $i=e1$.

Since in $\mathcal{G}$, the neighbors of the item nodes include both user nodes and entity nodes, separating message aggregation for the neighbor nodes based on the node type is necessary to generate the item embedding representations. We finally formulate the embedding generation in LS-GCN-2 for the target item $i1$ as follows:
\begin{equation}\label{eq:i_embed_2}
{v^*}_{i1}^{(l)} =  \sum_{u \in N_{i1}^{(u)}\cup i1}{\frac{(s_{i1})^{\gamma}\left|\mathcal{I}\right|}{\sum_{i \in \mathcal{I}}{(s_i)^{\gamma}}} \cdot \frac{{v^*}_u^{(l-1)}}{\left|N_{i1}\right|^{0.5}\left|N_u\right|^{0.5-\alpha}} } + \sum_{e \in N_{i1}^{(e)}\cup i1}{\frac{(s_{i1})^{\gamma}\left|\mathcal{I}\right|}{\sum_{i \in \mathcal{I}}{(s_i)^{\gamma}}} \cdot \frac{{v^*}_e^{(l-1)}}{\left|N_{i1}\right|^{0.5}\left|N_e\right|^{0.5-\alpha}} },
\end{equation}
where $N_{i1}^{(u)}$ and $N_{i1}^{(e)}$ denote the user and entity neighbor nodes of item $i1$ in $\mathcal{G}$, respectively, and ${v^*}_e^{(0)}$ is the initialized embedding of the entity $e$ in Equation \ref{eq:initial}, which is equivalent to ${v}_e^{(0)}$.

\subsection{Model Prediction \& Optimization Layer}\label{subsec:pred}
\subsubsection{\textbf{Prediction Function}}\label{subsubsec:pred_func}
In GCN, the node embeddings obtained at layer $l$ already preserve the information from all the previous layers. Therefore we choose the outputs of the last layer as the final representations of all the nodes. Considering an $L$-layer GCN here, for a target user $u$ and item $i$, we apply the inner product to calculate the user's preference score for the item as follows:
\begin{equation}\label{eq:pred}
{\hat{y}}_{ui} = (v_u^{(L)})^T\cdot v_i^{(L)} + ({v^*}_u^{(L)})^T\cdot {v^*}_i^{(L)}.
\end{equation}
\subsubsection{\textbf{Objective Function}}\label{subsubsec:obj_func}
In order to optimize the MEGCF, we select the BPR loss \cite{bprloss} as a base objective function, which is used in a wide range of recommendation methods. The core idea of BPR loss is that the preference score between the observed user-item pair is higher than that of the unobserved one.

Firstly, in order to ensure the learning of high-order CF signal, we construct the BPR loss using the embeddings output from LS-GCN-1, that is, the embeddings in Equations \ref{eq:i_embed_1} and \ref{eq:u_embed_1}.
\begin{equation}\label{eq:loss1}
\mathcal{L}_1 = \sum_{(u,i,j)\in \mathcal{O}}-ln\sigma({[v_u^{(L)}]}^T \cdot v_i^{(L)} - {[v_u^{(L)}]}^T \cdot v_j^{(L)}) + {\lambda}_1 \cdot ||\mathcal{H}_1||_2^2,
\end{equation}
where $\mathcal{O}=\{(u,i,j)|(u,i)\in \mathcal{R}^+, (u,j)\notin \mathcal{R}^+\}$ is the full training data, $\mathcal{R}^+$ denotes the full observed user-item interactions in $\mathcal{G}_1$, $\sigma(\cdot)$ is a sigmoid function, $\mathcal{H}_1=\{{\mathcal{V}_u}^{(L)},{\mathcal{V}_i}^{(L)}\}$ denotes the trainable parameters in this step, ${\mathcal{V}_u}^{(L)}=\{{{v}_{u_1}^{(L)}},...,{{v}_{u_{|\mathcal{U}|}}^{(L)}}\}$ and  ${\mathcal{V}_i}^{(L)}=\{{{v}_{i_1}^{(L)}},...,{{v}_{i_{|\mathcal{I}|}}^{(L)}}\}$ are the user and item embeddings obtained at $L$-th layer in LS-GCN-1, respectively,
and ${\lambda}_1$ is the coefficient of $L_2$ regularization for $\mathcal{H}_1$.

Afterwards, in order to facilitate the capture of high-order multimodal semantic correlations, we utilize the embeddings output from LS-GCN-2 ($i.e.$, the embeddings in Equations \ref{eq:u_embed_2} and \ref{eq:i_embed_2}) to construct the corresponding BPR loss:
\begin{equation}\label{eq:loss2}
\mathcal{L}_2 = \sum_{(u,i,j)\in \mathcal{O}}-ln\sigma({[ {v^*}_u^{(L)}]}^T\cdot{{v^*}_i^{(L)}} - {[ {v^*}_u^{(L)}]}^T\cdot{ {v^*}_j^{(L)}})
+{\lambda}_2 \cdot ||\mathcal{H}_2||_2^2,
\end{equation}
where $\mathcal{H}_2=\{{\mathcal{V^*}_u^{(L)}},{\mathcal{V^*}_i^{(L)}}\}$ denotes the trainable parameters in this step, ${\mathcal{V^*}_u}^{(L)}=\{{{v^*}_{u_1}^{(L)}},...,{{v^*}_{u_{|\mathcal{U}|}}^{(L)}}\}$ and ${\mathcal{V^*}_i}^{(L)}=\{{{v^*}_{i_1}^{(L)}},...,{{v^*}_{i_{|\mathcal{I}|}}^{(L)}}\}$ are the user and item embeddings obtained at $L$-th layer in LS-GCN-2, respectively,
and ${\lambda}_2$ is the coefficient of $L_2$ regularization for $\mathcal{H}_2$. 
It is worth mentioning that since the embeddings of users and items are incorporated with the information of multimodal entities through multiple message propagation, the optimization of this objective function can be considered as using the interactions to support the learning of user preferences over multimodal semantics.

Finally, we propose an objective function to jointly learn the Equations \ref{eq:loss1} and \ref{eq:loss2} as follows:
\begin{equation}\label{eq:loss}
\mathcal{L} = \mathcal{L}_1 +  \mathcal{L}_2.
\end{equation}
We adopt the mini-batch Adam optimizer \cite{adam} to minimize the loss in Equation \ref{eq:loss} and update the model parameters.

\subsection{Complexity Analysis of MEGCF}
Here we analyze the complexity of MEGCF. To the best of our knowledge, LightGCN is the most efficient GCN-based recommendation model. Therefore, we compare its complexity with that of MEGCF. As for the model size, the model parameters introduced by LightGCN are the initialized embeddings of users and items, while MEGCF additionally introduces the embeddings of entity nodes. In the following, we analyze the time complexity of MEGCF and LightGCN for the complete model training process. 

Assuming that the number of edges on the user-item interaction graph $\mathcal{G}_1$ and the multimodal interaction graph $\mathcal{G}$ are $|\mathcal{E}|$ and $|\mathcal{E}_m|$, respectively, $d$ denotes the embedding length, $s$ represents the number of training epochs, $B$ is the size of each training batch, and $L$ denotes the depth of GCN layers. Their computational complexity comes mainly from two parts: (1) the graph convolution process, and (2) the calculation of BPR loss.
\begin{itemize}[leftmargin=*]
	\item For the graph convolution process, the time complexities of MEGCF to perform one graph convolution on $\mathcal{G}_1$ and $\mathcal{G}$ are $O(2|\mathcal{E}|)$ and $O(2|\mathcal{E}_m|)$, respectively. Thus, its complexity in the whole training process is $O(2(|\mathcal{E}|+|\mathcal{E}_m|)Lds\frac{|\mathcal{E}|}{B})$, while the complexity of LightGCN is $O(2|\mathcal{E}|Lds\frac{|\mathcal{E}|}{B})$.
	\item For the calculation of BPR loss, the scoring prediction is the core operation to be considered. Note that both MEGCF and LightGCN use a simple inner product as the prediction function and its complexity is $O(d)$. Thus, the time complexity for LightGCN at this part in the whole training process is $O(2|\mathcal{E}|ds)$. Since MEGCF needs to perform scoring predictions on $\mathcal{G}_1$ and $\mathcal{G}$, respectively ($cf.$ Equations \ref{eq:loss1} and \ref{eq:loss2}), the corresponding time complexity of MEGCF is twice that of LightGCN, $i.e.$, $O(4|\mathcal{E}|ds)$.
\end{itemize}
Therefore, the overall training complexity of the proposed MEGCF is close to $O(2(|\mathcal{E}|+|\mathcal{E}_m|)Lds\frac{|\mathcal{E}|}{B} + 4|\mathcal{E}|ds)$, while the complexity of LightGCN is $O(2|\mathcal{E}|Lds\frac{|\mathcal{E}|}{B} + 2|\mathcal{E}|ds)$.

\section{EXPERIMENT}\label{sec:experiment}
In this section, we conduct experiments to evaluate our proposed MEGCF, and some ablation studies to verify the effectiveness of each component in MEGCF. We aim to answer three main research questions as follows:
\begin{itemize}[leftmargin=*]
	\item \textbf{RQ1:} How does our proposed MEGCF perform compared with the state-of-the-art baselines?
 	\item \textbf{RQ2:} Whether the components (modality-specific semantic correlation, symmetric linear GCN structure, sentiment-weighted neighbor aggregation, and joint loss function) in MEGCF are effective?
	\item \textbf{RQ3:} Whether the capture of multimodal semantic correlations is helpful for modeling modality-level item similarity and user preference?
\end{itemize}

\subsection{Experimental Settings}
\subsubsection{\textbf{Datasets.}}
Since our work aims to study multimodal data processing in recommendations, we choose two real-world datasets from \emph{Amazon.com} introduced by \cite{amazon-dataset}, \textbf{Beauty} and \textbf{Arts\_crafts\_and\_Sewing} (short for \textbf{Art})\footnote{http://deepyeti.ucsd.edu/jianmo/amazon/index.html}; both of them contain images, titles, and reviews. Besides, we select another fashion collocation dataset, \textbf{Taobao}\footnote{https://tianchi.aliyun.com/competition/entrance/231506/information}, which is published in the Tianchi competition. This dataset contains images and titles, except reviews, so the MEGCF run on this dataset is not equipped with the strategy of sentiment-weighted neighbor aggregation. To ensure the quality of these three datasets, we apply the 5-core setting, that is, retaining that all users and items have at least five interactions.

We present the details of these three datasets in Table \ref{tab:statistics}. Following the convention setting\cite{ncf,dicf,vbpr}, we apply the \emph{leave-one-out} evaluation\cite{bprloss} to randomly sample one item for each user to form the test set and another one to form the validation set, and the remaining interaction data to serve as the training set.
\renewcommand{\arraystretch}{1.4}
\renewcommand\tabcolsep{2.8 pt} 
\begin{table}
	\centering
	\fontsize{8.5}{8.5}\selectfont
	\caption{Statistics of the datasets, where \# VE and \# TE denote the number of visual and textual entities introduced in Section \ref{subsec:mmem}, respectively, and density is calculated by using ${\# Interaction}/{(\# User \times \# Item)}$.}
	\label{tab:statistics}
	\begin{tabular}{|c|c|c|c|c|c|c|}
		\hline
		\bf{Dataset}&\bf{\# User}&\bf{\# Item}&\bf{\# Interaction}&\bf{Density}&\bf{\# VE}&\bf{\# TE}\cr
		\hline
		{{Beauty}}&15,576&8,678&139,318&0.00103&1,080&11,450\cr
		\hline
		{Art}&25,165&9,324&201,427&0.00086&962&11,215\cr
		\hline
		{Taobao}&12,539&8,735&83,648&0.00076&1,127&8,476\cr
		\hline
	\end{tabular}
\end{table}

\subsubsection{\textbf{Evaluate Metrics.}} We select two protocols: \emph{Hit Ratio (HR)} and \emph{Normalized Discounted Cumulative Gain (NDCG)}, which are widely used in recent works \cite{ncf}\cite{dicf}\cite{apr} to evaluate model performance. Specifically, we compute the average $HR@k$ and $NDCG@k$ for each user in the test set. Note that for each user, we randomly sample 99 items from all items that the user has not interacted with as negative samples.

\subsubsection{\textbf{Baselines.}}\label{subsubsec:baselines}
To demonstrate the effectiveness of our proposed MEGCF, we compare it with the following baselines:
\begin{itemize}[leftmargin=*]
	\item \textbf{BPRMF}\cite{mf}: Matrix Factorization (MF) is a classical collaborative filtering method, which is widely used as a recommender baseline. BPRMF optimizes MF using BPR loss.
	\item \textbf{SVD++}\cite{svd++}: this is a variant of MF, which integrates the historical interactions into user embeddings. It can also be viewed as a one-layer linear GCN that only passes messages for user nodes. To ensure fairness, we employ BPR loss to optimize this baseline.
	\item \textbf{VBPR}\cite{vbpr}: this model incorporates visual features into the item representations and applies the MF framework to predict the preference scores of user-item pairs.
	\item \textbf{CKE}\cite{CKE-kdd2016}: This model incorporates visual features, textual features, and knowledge graph into the item representations, and uses MF as the overall framework. In the experiments, we consider only visual and textual features  since there is no available knowledge graph in the datasets.
	\item \textbf{NGCF} \cite{ngcf}: this model adopts a nonlinear GCN to iteratively perform message propagation on the user-item graph and concatenates the embedding obtained from each GCN layer as the final representations of the user and item nodes.
	\item \textbf{MMGCN} \cite{mmgcn}: this is a multimodal recommendation model, which considers features of visual, textual, and audio modalities. It also applies three nonlinear GCNs to perform message passing on the user-item graphs that hold data of different modalities, respectively, so as to learn fine-grained modality-specific user preferences. It finally fuses embeddings of different modalities as the final representations of users and items. In the experiment, we consider only visual and textual features due to the limitation of the datasets. 
	\item \textbf{LightGCN} \cite{light}: this is the state-of-the-art GCN-based CF model, which incorporates a linear GCN into CF scenarios  and uses the summation of the embeddings obtained at each layer as the final representation.
	\item \textbf{GRCN} \cite{grcn}: this is the state-of-the-art  multimodal recommendation method, whose main framework can be viewed as a linear GCN, where multimodal features of items are used to weight the neighbor aggregation, and finally the output of each graph convolution layer is summed and concatenated with the multimodal features as the final node representations.
\end{itemize}

\subsubsection{\textbf{Hyper-parameter Settings.}}
For all methods of comparison, we set the embedding size and batch size to 64 and 2048, respectively. We tune the learning rate in \{$10^{-4}$, $10^{-3}$, $10^{-2}$, $10^{-1}$, $1$\} and search the coefficient of $L_2$ regularization in \{$10^{-5}$, $10^{-4}$, $10^{-3}$, $10^{-2}$, $10^{-1}$, 1\}. For GCN-based methods, \textit{i.e.}, NGCF, MMGCN, LightGCN, GRCN, and our proposed MEGCF, we tune the number of graph convolution layers in \{1, 2, 3, 4, 5, 6\}. Besides, we use the Xavier initializer \cite{xavier} to achieve the embedding initialization for all models.

\subsection{Overall Comparison (RQ1)}

\renewcommand{\arraystretch}{1.3}
\renewcommand\tabcolsep{3 pt} 
\begin{table}
	\centering
	\fontsize{8.5}{8.5}\selectfont
	\caption{Overall performance comparison.}
	\label{tab:overall}
	\begin{tabular}{|c|c|ccc|ccc|ccc|}
		\hline
		\multirow{2}{*}{\textbf{Metric}} & \multirow{2}{*}{\textbf{Models}} & \multicolumn{3}{c|}{\textbf{Beauty}} & \multicolumn{3}{c|}{\textbf{Art}} & \multicolumn{3}{c|}{\textbf{Taobao}} \cr\cline{3-11}
         & & $k$=5    & $k$=10    & $k$=20    & $k$=5   & $k$=10   & $k$=20   & $k$=5    & $k$=10    & $k$=20   \cr\hline\hline

        \multirow{8}{*}{\bf{HR@$k$}}
        & {BPRMF} &{0.4274}&{0.5173}&{0.6231}&{0.6333}&{0.7052}&{0.7829}&{0.3215}&{0.4049}&{0.5155}\cr\cline{2-11}
        & {SVD++} &{0.4584}&{0.5520}&{0.6659}&{0.6530}&{0.7425}&{0.8285}&{0.3374}&{0.4293}&{0.5466}\cr\cline{2-11}
        & {VBPR} &{0.4722}&{0.5670}&{0.6665}&{0.6699}&{0.7464}&{0.8262}&{0.3464}&{0.4364}&{0.5512}\cr\cline{2-11}
        & {CKE} &{0.4810}&{0.5894}&{0.6950}&{0.6719}&{0.7632}&{0.8461}&{0.3560}&{0.4550}&{0.5789}\cr\cline{2-11}
        & {NGCF} &{0.4853}&{0.5820}&{0.6810}&{0.6742}&{0.7541}&{0.8287}&{0.3575}&{0.4593}&{0.5841}\cr\cline{2-11}
        & {MMGCN} &{0.4934}&{0.6067}&{0.7166}&{0.6769}&{0.7702}&{0.8546}&{0.3649}&{0.4695}&{0.5902}\cr\cline{2-11}
        & {LightGCN} &{0.5002}&{0.6063}&{0.7178}&{0.6814}&{0.7639}&{0.8329}&{0.3848}&{0.4893}&{0.6237}\cr\cline{2-11}
        & {GRCN} &{\underline{0.5087}}&{\underline{0.6204}}&{\underline{0.7241}}&{\underline{0.6905}}&{\underline{0.7743}}&{\underline{0.8532}}&{\underline{0.3865}}&{\underline{0.4996}}&{\underline{0.6375}}\cr\cline{2-11}
        & \bf{MEGCF} &\textbf{0.5439}&\textbf{0.6464}&\textbf{0.7448}&\textbf{0.7116}&\textbf{0.7902}&\textbf{0.8651}&\textbf{0.4045}&\textbf{0.5212}&\textbf{0.6516}\cr\cline{2-11}
        & $\%Improv.$ &{6.92\%}&{4.19\%}&{2.86\%}&{3.06\%}&{2.05\%}&{1.39\%}&{4.65\%}&{4.32\%}&{2.21\%}\cr\cline{2-11}\hline\hline

        \multirow{8}{*}{\bf{NDCG@$k$}}
        & {BPRMF} &{0.3343}&{0.3634}&{0.3900}&{0.5597}&{0.5829}&{0.6025}&{0.2465}&{0.2733}&{0.3011}\cr\cline{2-11}
        & {SVD++} &{0.3592}&{0.3895}&{0.4157}&{0.5627}&{0.5916}&{0.6134}&{0.2523}&{0.2819}&{0.3114}\cr\cline{2-11}
        & {VBPR} &{0.3665}&{0.3973}&{0.4224}&{0.5830}&{0.6078}&{0.6280}&{0.2639}&{0.2928}&{0.3216} \cr\cline{2-11}
        & {CKE} &{0.3650}&{0.4002}&{0.4269}&{0.5739}&{0.6030}&{0.6245}&{0.2622}&{0.2941}&{0.3253}\cr\cline{2-11}
        & {NGCF} &{0.3776}&{0.4089}&{0.4339}&{0.5882}&{0.6141}&{0.6330}&{0.2658}&{0.2986}& {0.3301}\cr\cline{2-11}
        & {MMGCN} &{0.3714}&{0.4081}&{0.4359}&{0.5643}&{0.5945}&{0.6159}&{0.2709}&{0.3047}& {0.3351}\cr\cline{2-11}
        & {LightGCN}&{0.3807}&{0.4152}&{0.4435}&{0.5886}&{0.6153}&{0.6340}&{0.2840}&{0.3176}&{0.3515}\cr\cline{2-11}
        & {GRCN} &{\underline{0.3910}}&{\underline{0.4272}}&{\underline{0.4533}}&{\underline{0.5937}}&{\underline{0.6208}}&{\underline{0.6407}}&{\underline{0.2861}}&{\underline{0.3225}}& {\underline{0.3573}}\cr\cline{2-11}
        & \bf{MEGCF} &\textbf{0.4257}&\textbf{0.4590}&\textbf{0.4838} &\textbf{0.6144}&\textbf{0.6398}&\textbf{0.6588}&\textbf{0.3020}&\textbf{0.3397}&\textbf{0.3726} \cr\cline{2-11}
         
        & $\%Improv.$ &{8.87\%}&{7.44\%}&{6.73\%}&{3.49\%}&{3.06\%}&{2.83\%}&{5.56\%}&{5.33\%} &{4.28\%} \cr\hline
\end{tabular}
\end{table}

\begin{figure}
	\centering
	\includegraphics[height=7.94cm, width=13cm]{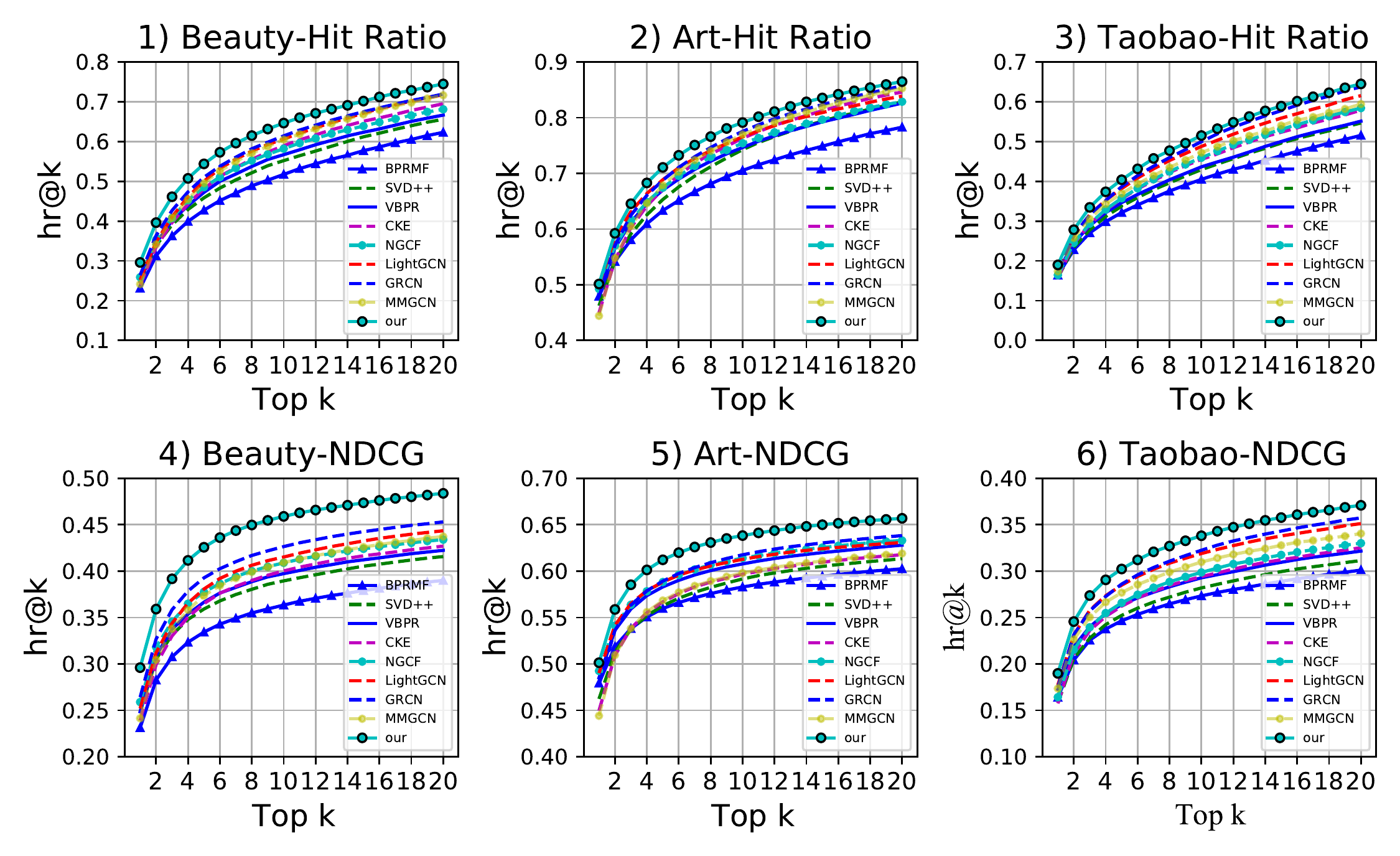}
	\caption{Performance comparison of all the models over the different $k$ on the three datasets.}
	\label{fig:over_all}
\end{figure}

To evaluate our proposed MEGCF, we compare it with traditional CF models (BPRMF and SVD++), GCN-based CF models (NGCF and LightGCN), and multimodal-based models (VBPR, CKE, MMGCN, and GRCN).
Table \ref{tab:overall} and Figure \ref{fig:over_all} report the performance of all the models. We obtain the following findings:
\begin{itemize}[leftmargin=*]
	\item BPRMF performs the weakest in all the cases, which indicates that the use of simple inner product strongly depends on sufficient interaction and makes it difficult to model complex interaction connectivity in sparse scenarios. SVD++ outperforms BPRMF on the three datasets, which demonstrates that explicitly incorporating historical interactions into the user embedding is helpful for modeling user preferences.
	\item The GCN-based methods (NGCF, MMGCN, LightGCN, GRCN) consistently outperform BPRMF and SVD++, which demonstrates the effectiveness of explicitly capturing high-order CF signals. In addition, LightGCN achieves a significant improvement over NGCF on the three datasets, which is due to the fact that the linear GCN used by LightGCN is more suitable for capturing high-order CF signals than the nonlinear GCN used by NGCF (evidence is also given in \cite{light}). 
	\item The multimodal baselines (VBPR, CKE, and MMGCN) always outperform SVD++, which demonstrates the effectiveness of modeling modality-level user preferences in solving sparsity problems. CKE slightly outperforms VBPR \textit{w.r.t.} HR in most cases, which is owning to the additional incorporation of deep textual features in CKE. However, CKE is often weaker than VBPR \textit{w.r.t.} NDCG especially on the Art dataset, which is attributed to the fact that the textual deep features that CKE additionally incorporates contain too much noise unrelated to user preferences, thus contaminating the embedding representation and weakening the modeling of ranking preferences.
	\item MMGCN outperforms NGCF in most cases, which indicates that incorporating multimodal features while capturing high-order CF signals can further improve the model performance. Unexpectedly, in terms of NDCG on the Art dataset, MMGCN is weaker than NGCF, CKE, and VBPR. This may be due to the fact that MMGCN extends multimodal features to a higher order through the user-item interaction graph, which in turn amplifies the preference-independent information and intensifies the contamination of embeddings, thereby weakening the modeling of user preferences in terms of ranking. The results of GRCN confirm this analysis, since GRCN, combining higher-order CF signal and lower-order multimodal features, consistently achieves a high performance $w.r.t.$ NDCG.
	\item MEGCF achieves the optimal performance on the three datasets, which demonstrates the importance of simultaneously mining high-order multimodal semantic correlation and CF signal. In particular, MEGCF achieves maximum improvements of 8.87\% and average improvements of 4.40\% compared with the strongest baseline GRCN (the method with an underline). 
	{An important observation is that MEGCF achieves more remarkable improvements on NDCG than on HR, which may be attributed to the fact that the multimodal semantic correlation captured in MEGCF is more advantageous for ranking preference modeling. In-depth experiments in Section \ref{subsubsec:mm} further validate this analysis, as after eliminating all the multimodal information ($i.e.$, the model variant \textbf{w/o V\&T}), the model performance decreases more on NDCG than on HR.}
	It is worth mentioning that the multimodal recommender baselines (VBPR, CKE, and MMGCN) perform poorly on NDCG, while MEGCF achieves a greater improvement on NDCG. This indicates that for alleviating the embedding contamination, mining semantic correlation from multimodal content ($i.e.$, the strategy in MEGCF) is more effective than incorporating complete multimodal deep features.

\end{itemize}

\subsection{Study of MEGCF (RQ2)}
In this section, we aim to investigate the effectiveness of all the components in MEGCF. Specifically, we first study the impact of modality-specific semantic correlation on model performance. Then, we further decompose the symmetric GCN modules in MEGCF and study the gains they bring to the model. Next, we investigate the effectiveness of the proposed joint loss function. After that, we compare the impacts of sentiment weighting strategy on different GCN-based methods. Finally, we assess the influence of the number of graph convolution layers in MEGCF and other GCN-based baselines.
\renewcommand{\arraystretch}{1.4}
\renewcommand\tabcolsep{2.8 pt} 
\begin{table}
	\centering
	\fontsize{8.5}{8.5}\selectfont
	\caption{Ablation study of components in MEGCF.}
	\label{tab:ablation}
	\begin{tabular}{|c|cc|cc|cc|cc|cc|cc|}
		\hline
		\multirow{3}{*}{\bf{Methods}}
		&\multicolumn{4}{c|}{\bf{Beauty}}&\multicolumn{4}{c|}{\bf{Art}}&\multicolumn{4}{c|}{\bf{Taobao}}\cr\cline{2-13}
		&\multicolumn{2}{c|}{\bf{HR@$k$}}&\multicolumn{2}{c|}{\bf{NDCG@$k$}}&\multicolumn{2}{c|}{\bf{HR@$k$}}&\multicolumn{2}{c|}{\bf{NDCG@$k$}}&\multicolumn{2}{c|}{\bf{HR@$k$}}&\multicolumn{2}{c|}{\bf{NDCG@$k$}}\cr\cline{2-13}
		&{$k$=10}&{$k$=20}&{$k$=10}&{$k$=20}&{$k$=10}&{$k$=20}&{$k$=10}&{$k$=20}&{$k$=10}&{$k$=20}&{$k$=10}&{$k$=20}\cr\hline\hline
		{w/o V\&T}&0.6047&0.7105&0.4124&0.4391&0.7477&0.8250&0.6131&0.6327&0.4978&0.6329&0.3209&0.3549\cr
	{w/o V}&0.6226&0.7271&0.4326&0.4589&0.7716&0.8466&0.6201&0.6391&0.5076&0.6362&0.3278&0.3603\cr
	{w/o T}&0.6150&0.7118&0.4328&0.4574&0.7642&0.8421&0.6185&0.6383&0.5079&0.6316&0.3304&0.3617\cr
		\hline
		{MEGCF$_{g1}$}&0.6026&0.7093&0.4145&0.4415&0.7542&0.8314&0.6145&0.6340&0.4990&0.6244&0.3241&0.3557\cr
		{MEGCF$_{g2}$}&0.6156&0.7345&0.4100&0.4400&0.7654&0.8448&0.6103&0.6304&0.4849&0.6217&0.3196&0.3540\cr
		\hline
		 {w/o $\mathcal{L}_1$}&0.5902&0.7171&0.3609&0.3930&0.7411&0.8455&0.5239&0.5504&0.4751&0.6355&0.2625&0.3030\cr
		 {w/o $\mathcal{L}_2$}&0.6161&0.7214&0.4255&0.4522&0.7355&0.8139&0.5975&0.6173&0.5047&0.6387&0.3250&0.3587\cr
		\hline
		{w/o PN}&{0.6359}&{0.7358}&{0.4509}&{0.4763}&{0.7809}&{0.8557}&{0.6297}&{0.6493}&{0.5109}&{0.6404}&{0.3329}&{0.3665}\cr
		\hline
		MEGCF&0.6464&0.7448&0.4590&0.4838&0.7912&0.8648&0.6383&0.6569&0.5212&0.6516&0.3397&0.3726\cr\hline
	\end{tabular}
\end{table}
\begin{figure}
	\centering
	\includegraphics[height=8.56cm, width=14cm]{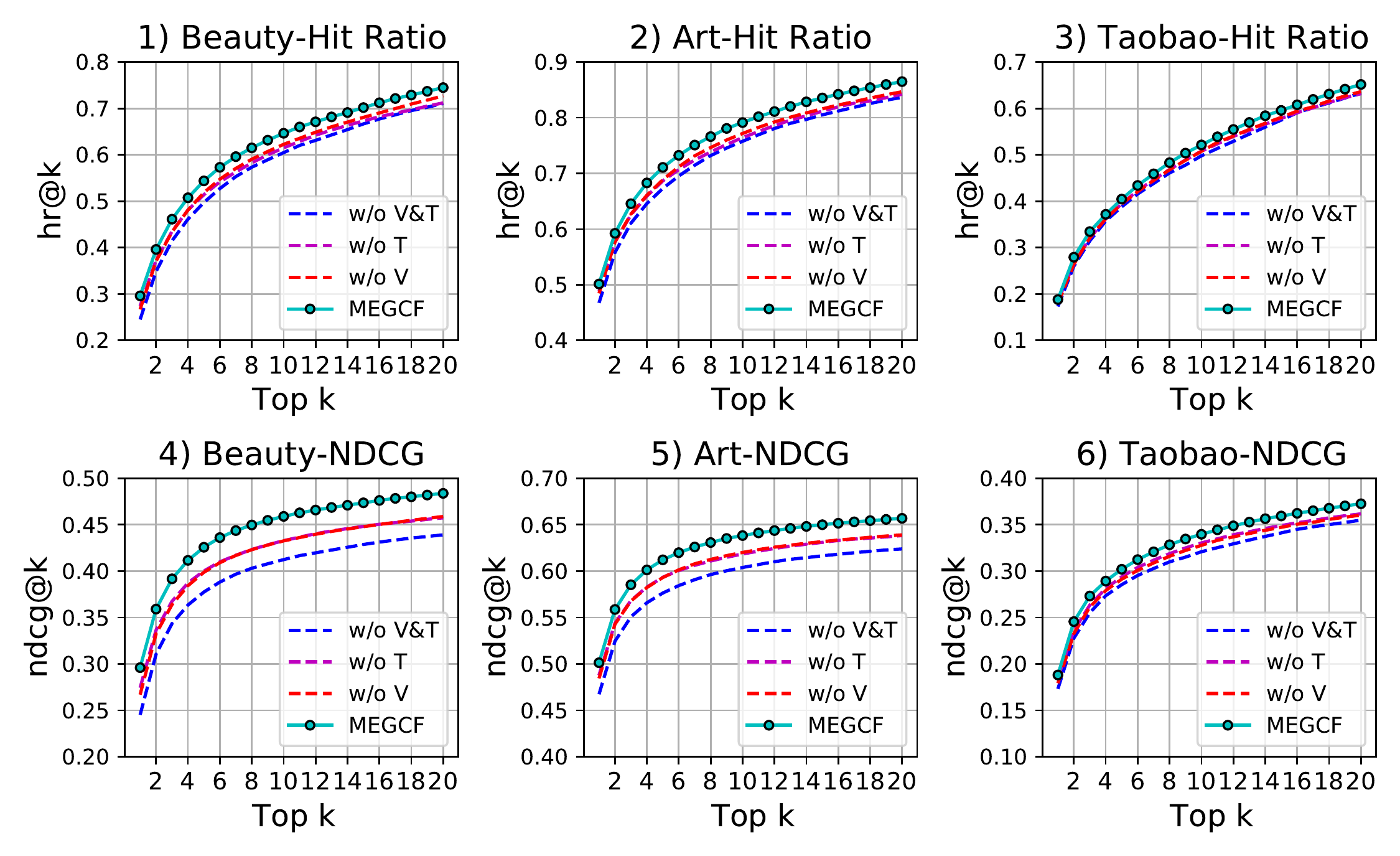}
	\caption{Effect of modality-specific semantic correlation on MEGCF.}
	\label{fig:3-3-1}
\end{figure}

\subsubsection{\textbf{Is modality-specific semantic correlation helpful?}}\label{subsubsec:mm}
MEGCF mines semantic entities from visual and textual modalities, respectively, and captures high-order multimodal semantic correlation using user-item interactions and these semantic entities. In order to study the impact of semantic correlation of different modalities on the model performance, we set up the following variants of MEGCF:
\begin{itemize}[leftmargin=*]
    \item \textbf{{w/o V}}: this model is obtained by removing the visual semantic entities from MEGCF.
    \item \textbf{{w/o T}}: this model is obtained by removing the textual semantic entities from MEGCF.
    \item \textbf{{w/o V\&T}}: this model removes both textual and visual semantic entities, which is equivalent to the symmetric GCN module in MEGCF capturing only the high-order CF signals. 
\end{itemize}
We conduct ablation experiments on the three datasets. Figure \ref{fig:3-3-1} and Table \ref{tab:ablation} record the model performance of the three variants and MEGCF \textit{w.r.t.} HR@$k$ and NDCG@$k$. We find that the curve of \textbf{{w/o V\&T}} always lies at the bottom, $i.e.$, the model performs worst when the multimodal semantic entities are not used, which indicates the importance of capturing multimodal semantic correlations. 
Furthermore, the curves of \textbf{w/o V} and \textbf{w/o T} considering visual or textual information alone are consistently close to each other, $i.e.$, the capture of visual and textual semantic correlations provides a similar gains in model performance. This may be due to the fact that in the recommendation scenarios for these three datasets (E-commerce), the visual and textual contents of the item contribute similarly to  triggering the user's interaction behavior ($i.e.$, whether the user will like the item or not).  
MEGCF consistently outperforms \textbf{w/o V} and \textbf{w/o T}, which indicates that the visual and textual features are significantly different in triggering user interest, and thus simultaneously modeling semantic correlation on multiple modalities can better model user preferences.

\subsubsection{\textbf{Is symmetric linear GCN block helpful?}}\label{subsubsec:g}

In section \ref{subsec:slgcn}, we propose two linear GCN, LS-GCN-1 and LS-GCN-2 (respectively denoted by g1 and g2). They can capture high-order CF signals and high-order multimodal semantic correlations, respectively. In order to investigate the effect of them, we set up the following variants of MEGCF:
\begin{figure}
	\centering
	\includegraphics[height=8.56cm, width=14cm]{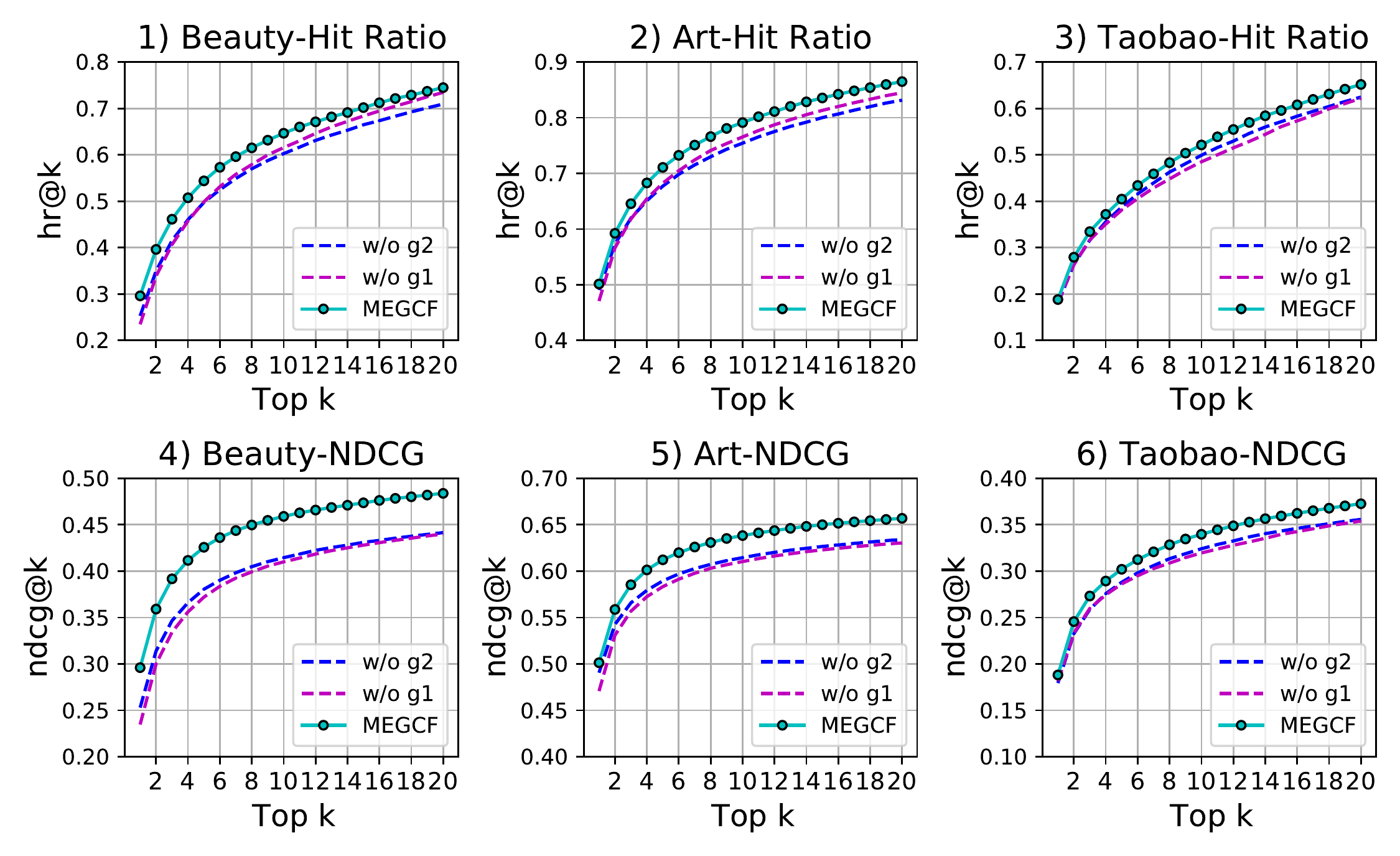}
	\caption{Effect of symmetric linear GCN module on MEGCF.}
	\label{fig:3-3-2}
\end{figure}
\begin{itemize}[leftmargin=*]
    \item {\textbf{w/o g2}}: this model retains only g1 in the symmetric linear GCN to generate the final embedding representations of the nodes.
    \item {\textbf{w/o g1}}: this model uses g2 rather than the symmetric linear GCN module for node embedding generation. 
    \item {\textbf{w/o PN}: this variant uses the classical Graph Laplacian Norm rather than the popularity-aware norm (short for PN, $cf.$ Equation \ref{eq:i_embed_1}).} 
\end{itemize}
We conduct experiments on these two variants and MEGCF using the three datasets. Figure \ref{fig:3-3-2} shows the top-$k$ recommendation performance of {\textbf{w/o g2}}, {\textbf{w/o g1}}, and MEGCF, and Table \ref{tab:ablation} records the specific performance of these methods. We have the following findings:

\begin{itemize}[leftmargin=*]	
	\item MEGCF consistently outperforms \textbf{w/o g2} and \textbf{w/o g1}, which indicates that there are significant differences between the CF signal and multimodal correlation captured by g1 and g2, respectively. Therefore, MEGCF combined with g1 and g2 can achieve complementarity between CF signal and multimodal correlation, and it further enhance the model performance.
	\item \textbf{w/o g1} is slightly weaker than \textbf{w/o g2} in most cases, which may be due to the difference between the captured CF signals in g1 and g2. More precisely, although g2 can be roughly considered as a simultaneous capture of the CF signal and semantic correlation,	the CF signal in g2 is mainly used to assist in the capture of semantic correlation. Therefore, it will inevitably incorporate preference-independent multimodal noise, which makes it less pure than the CF signal captured alone in g1. This leads to the result that\textbf{ w/o g1} is weaker than \textbf{w/o g2}.
	\item It can be seen from Table \ref{tab:ablation} that MEGCF is stronger than {\bf{w/o PN}} on the three datasets, which indicates the effectiveness of the popularity-aware norm in symmetric linear GCN module. Note that this result is consistent with that in \cite{jmpgcf}.
\end{itemize}

\subsubsection{\textbf{Is joint loss function helpful?}}
\begin{figure}
	\centering
	\includegraphics[height=8.56cm, width=14cm]{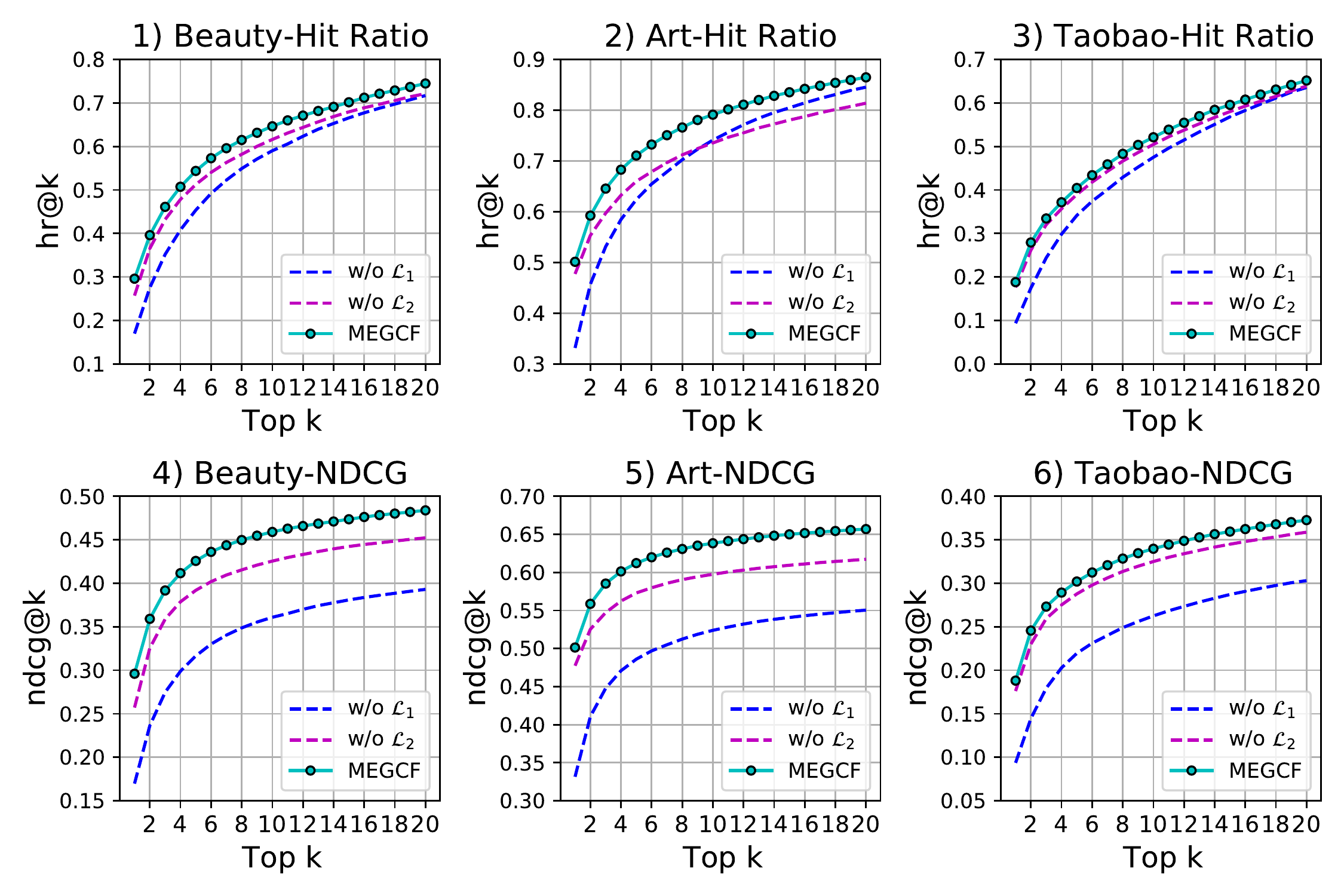}
	\caption{Effect of joint loss function on MEGCF.}
	\label{fig:3-3-3}
\end{figure}
In section \ref{subsubsec:obj_func}, we set the corresponding loss functions $\mathcal{L}_1$ and $\mathcal{L}_2$ for optimizing the model learning for CF signals and multimodal semantic correlations, respectively, while the final joint loss function is their simple summation. In order to investigate the effectiveness of the joint loss function in MEGCF, we set up the following model variants:
\begin{itemize}[leftmargin=*]
    \item \textbf{{w/o $\mathcal{L}_1$}}: this variant removes $\mathcal{L}_1$ from the final loss function of MEGCF. That is, its optimization goal is to capture multimodal semantic correlations.
    \item \textbf{{w/o $\mathcal{L}_2$}}: this variant removes $\mathcal{L}_2$ from the final loss function of MEGCF, which indicates that its optimization goal is to capture high-order CF signals, while the modeling of multimodal semantic correlations is weaker.

\end{itemize}
Figure \ref{fig:3-3-3} and Table \ref{tab:ablation} document the trend plots and the specific values of model performance for these model variants on the three datasets, respectively. \textbf{{w/o $\mathcal{L}_2$}} outperforms \textbf{{w/o $\mathcal{L}_1$}} in most cases, which indicates that simply optimizing the capture of multimodal semantic correlations is insufficient. In addition, MEGCF achieves a better performance than \textbf{{w/o $\mathcal{L}_1$}} and \textbf{{w/o $\mathcal{L}_2$}} on the three datasets, which demonstrates the effectiveness of the strategy of using a joint loss function to simultaneously optimize both the high-order CF signals and multimodal semantic correlations.

\subsubsection{\textbf{Is sentiment-weighted neighbor aggregation helpful?}}\label{subsubsec:ex:sentiment}

In MEGCF, we extract the sentiment information of users from the item's reviews and then use it to assign weights to this item node, which allows the model to distinguish the importance of different neighbors during the graph convolution process. In order to investigate whether the sentiment-weighted neighbor aggregation approach can improve the performance of MEGCF, and also to explore whether this approach is equally effective for other GCN-based recommender methods, we set up the following model variants:
\begin{itemize}[leftmargin=*]
    \item \textbf{NGCF}, \textbf{LightGCN}, and \textbf{MMGCN}: the GCN-based baselines introduced in Section \ref{subsubsec:baselines}.
    \item \textbf{NGCF$_s$}, \textbf{LightGCN$_s$}, and \textbf{MMGCN$_s$}: these model variants are obtained by applying the sentiment-weighted neighbor aggregation to NGCF, LightGCN, and MMGCN, respectively.
    \item \textbf{{w/o S}}: this variant is obtained by removing the sentiment weighting strategy from MEGCF.
\end{itemize}

\renewcommand{\arraystretch}{1.4}
\renewcommand\tabcolsep{2.8 pt} 
\begin{table}
	\centering
	\fontsize{8.5}{8.5}\selectfont
	\caption{Effect of the sentiment-weighted neighbor aggregation on GCN-based methods.}
	\label{tab:3-3-4}
	\begin{tabular}{|c|ccc|ccc|ccc|ccc|}
		\hline
		\multirow{3}{*}{\bf{Methods}}
		&\multicolumn{6}{c|}{\bf{Beauty}}&\multicolumn{6}{c|}{\bf{Art}}\cr\cline{2-13}
		&\multicolumn{3}{c|}{\bf{HR@$k$}}&\multicolumn{3}{c|}{\bf{NDCG@$k$}}&\multicolumn{3}{c|}{\bf{HR@$k$}}&\multicolumn{3}{c|}{\bf{NDCG@$k$}}\cr\cline{2-13}
		&{$k$=5}&{$k$=10}&{$k$=20}&{$k$=5}&{$k$=10}&{$k$=20}&{$k$=5}&{$k$=10}&{$k$=20}&{$k$=5}&{$k$=10}&{$k$=20}\cr\hline\hline
		NGCF&0.4853&0.5820&0.6810&0.3764&0.4089&0.4339&0.6742&0.7541&0.8287&0.5882&0.6141&0.6330\cr
		NGCF$_s$&0.4878&0.5828&0.6821&0.3776&0.4114&0.4354&0.6784&0.7565&0.8295&0.5911&0.6167&0.6362\cr
		$\%Improv.$&0.52\%&0.14\%&0.16\%&0.32\%&0.61\%&0.35\%&0.63\%&0.32\%&0.10\%&0.49\%&0.42\%&0.51\%\cr\hline
		MMGCN&0.4934&0.6067&0.7166&0.3714&0.4081&0.4359&0.6736&0.7681&0.8526&0.5627&0.5933&0.6148\cr
		MMGCN$_s$&0.4981&0.6092&0.7176&0.3746&0.4117&0.4379&0.6756&0.7695&0.8531&0.5685&0.5988&0.6180\cr
		$\%Improv.$&0.95\%&0.41\%&0.14\%&0.86\%&0.88\%&0.46\%&0.30\%&0.18\%&0.06\%&1.03\%&0.93\%&0.52\%\cr\hline
		LightGCN&0.5002&0.6063&0.7178&0.3807&0.4152&0.4435&0.6814&0.7639&0.8376&0.5886&0.6153&0.6340\cr
		LightGCN$_s$&0.5040&0.6100&0.7207&0.3838&0.4181&0.4460&0.6834&0.7647&0.8383&0.5897&0.6166&0.6350\cr
		$\%Improv.$&0.76\%&0.78\%&0.14\%&1.40\%&1.13\%&0.54\%&0.29\%&0.10\%&0.08\%&0.19\%&0.21\%&0.16\%\cr\hline
		{w/o S}&0.5332&0.6373&0.7404&0.4135&0.4472&0.4733&0.7052&0.7854&0.8587&0.6075&0.6335&0.6521\cr
		MEGCF&0.5439&0.6464&0.7448&0.4257&0.4590&0.4838&0.7116&0.7902&0.8651&0.6144&0.6398&0.6588\cr
		$\%Improv.$&2.01\%&1.43\%&0.59\%&2.95\%&2.64\%&2.22\%&0.91\%&0.61\%&0.75\%&1.14\%&0.99\%&1.03\%\cr\hline
	\end{tabular}
\end{table}
Table \ref{tab:3-3-4} documents the specific performance of these methods. We have the following findings:
\begin{itemize}[leftmargin=*]
	\item NGCF$_s$, MMGCN$_s$, LightGCN$_s$, and MEGCF generally outperform NGCF, MMGCN, LightGCN, and {w/o S}, respectively. This demonstrates that the sentiment-weighted approach not only brings gains for MEGCF, but it is also effective for other GCN-based methods.
	\item The sentiment weighting strategy achieves more significant improvement on MEGCF, compared with other GCN-based methods, which may be attributed to the symmetric graph convolution structure of MEGCF. Specifically, MEGCF is equipped with two different linear GCN that both benefit from the sentiment-weighted neighbor aggregation, which results in more improvement on MEGCF.
	\item An important phenomenon is that the improvement of these methods on NDCG is generally higher than that on HR. In addition, the smaller the size of the recommendation list ($i.e.$, $k$), the higher the improvement of these methods. All these results reflect the outperformance of sentiment-weighted neighbor aggregation methods in ranking preference modeling. Specifically, these GCN-based methods with sentiment weighting will tend to rank items of interest to users more towards the top of the top-$k$ recommendation list. Therefore, they lead to more improvement when $k$ is smaller or when NDCG is computed.
	\item The sentiment-weighted neighbor aggregation method has a significantly higher improvement on the Beauty dataset (almost 1.97\%) than on the Art dataset (almost 0.91\%), which is coherent with the results of overall comparison in Table \ref{tab:overall}. This is probably because in the recommendation scenario corresponding to the Art dataset, using only interaction data can achieve remarkable recommendation results (HR@5 > 0.7), which means that the room for further improvement will be more limited.
\end{itemize}

\subsubsection{\textbf{Is sentiment weighting superior to self-attention weighting ?}}\label{subsubsec:ex:sentiment-gat}
\begin{figure}
	\centering
	\includegraphics[height=4.67cm, width=14cm]{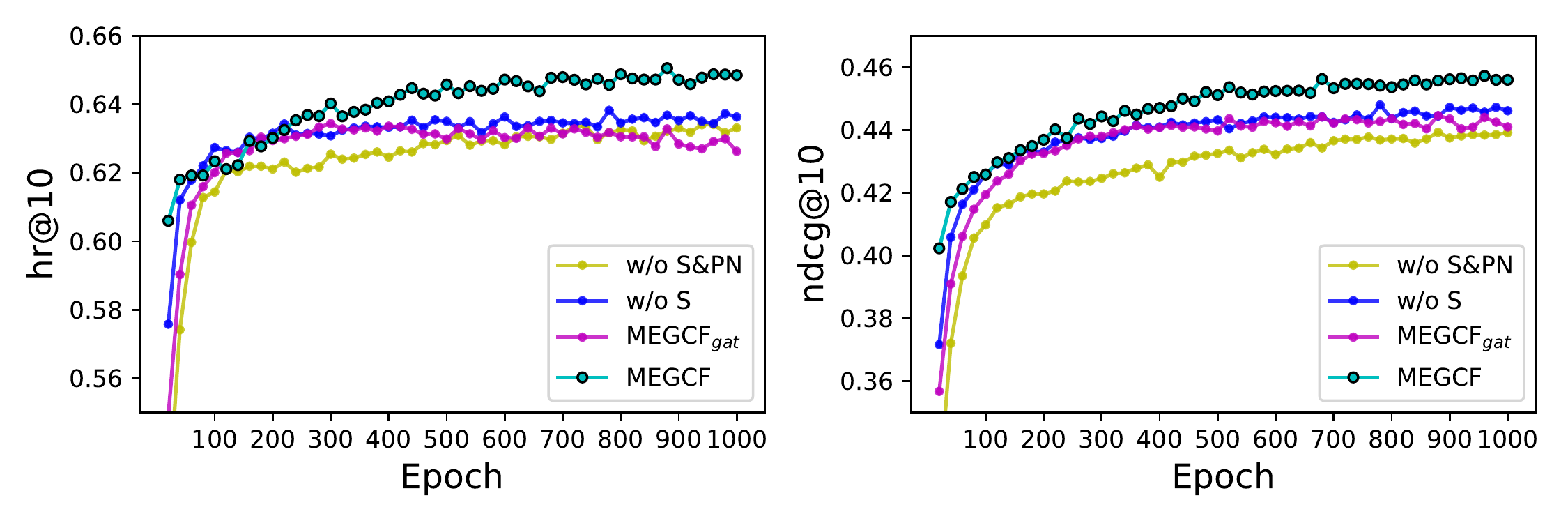}
	\caption{Performance trends on different training epochs on the Beauty dataset.}
	\label{fig:megcf-gat}
\end{figure}
In this part, in order to compare the influences of the review-based sentiment weighting strategy and the self-attention weighting of GAT\cite{gat} on the neighbor aggregation in GCN, we set up the following model variants:
\begin{itemize}[leftmargin=*]
	\item \textbf{w/o S\&PN}: this variant removes both the sentiment weighting strategy and the populairty-aware norm (short for PN) from MEGCF.
	\item \textbf{MEGCF$_{gat}$}: this variant uses the self-attention networks in GAT to weight the neighbor aggregation in MEGCF, rather than the sentiment weighting strategy.
\end{itemize}
Figure \ref{fig:megcf-gat} records the performance trends of these methods on the training epochs of 0-1000. we have the following findings:
\begin{itemize}[leftmargin=*]
	\item MEGCF is significantly stronger than MEGCF$_{gat}$, which indicates that sentiment weighting outperforms self-attention weighting in GAT for user preference modeling. Furthermore, we attribute this outperformance to the utilization of both interaction data and reviews in the sentiment weighting strategy, while the weights in MEGCF$_{gat}$ are only learned from interaction data.
	\item The curve of MEGCF$_{gat}$ and that of \textbf{w/o S} are close. This may be due to the fact that self-attention weighting in MEGCF$_{gat}$ and popularity-aware norm (PN) in \textbf{w/o S} exert similar effects on the model, both amplifying the CF signal ($i.e.$, assigning higher weights to nodes with more interactions). We leave this phenomenon for future studies. In the late training period (when the training epoch is greater than 700), the performance of MEGCF$_{gat}$ starts to slightly decrease, probably because the self-attention networks introduce more learnable parameters (weight matrices), which not only increases the model complexity but also makes the model more prone to the risk of overfitting.
	\item MEGCF$_{gat}$ significantly outperforms \textbf{w/o S\&PN} in the early stage of training, which demonstrates the effectiveness of the self-attention weighting strategy. However, in the late stage of training, \textbf{w/o S\&PN} performs very close to MEGCF$_{gat}$, which may be because the lightweight GCN used in these model variants has a strong ability to mine the interaction relations, and therefore the performance gain brought by the self-attention network in MEGCF$_{gat}$ can be replaced by adequate model training in \textbf{w/o S\&PN}.
\end{itemize}

\subsubsection{\textbf{Effect of the number of graph convolution layers.}}
\begin{figure}
	\centering
	\includegraphics[height=9.33cm, width=14cm]{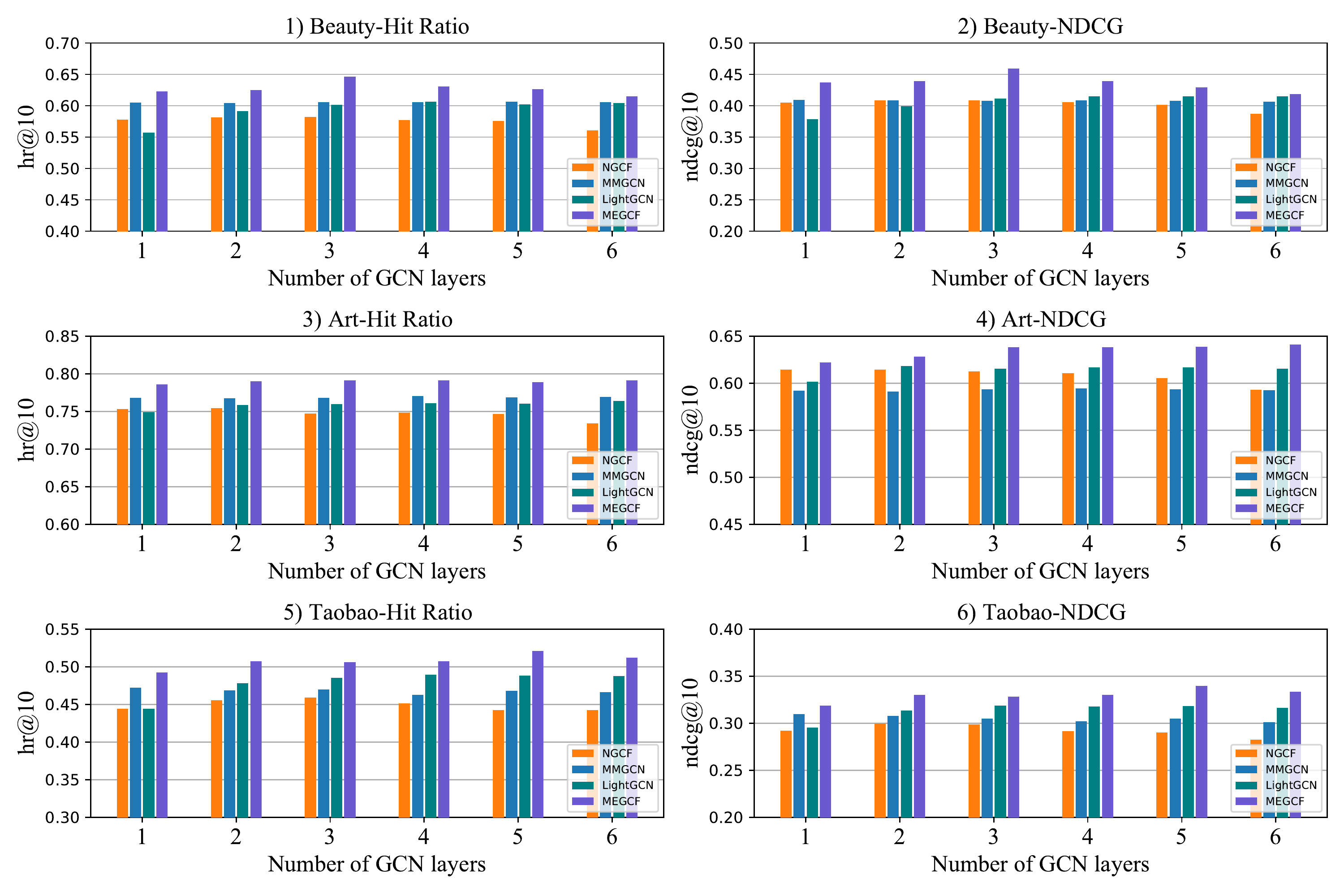}
	\caption{Performance of GCN-based methods (NGCF, LightGCN, MMGCN, and MEGCF) \textit{w.r.t.} the number of GCN layers on the three datasets.}
	\label{fig:3-3-5}
\end{figure}

In order to better investigate the ability of MEGCF to capture multimodal correlation and CF signal, we compare the model performance of MEGCF and other GCN-based baselines (NGCF, MMGCN, and LightGCN) \textit{w.r.t.} different graph convolution layers (we search the number of layers in \{1,2,3,4,5,6\}), as shown in Figure \ref{fig:3-3-5}. We have the following findings:
\begin{itemize}[leftmargin=*]
    \item Generally speaking, MEGCF achieves the optimal performance for all layer settings on the three datasets, which further demonstrates the effectiveness of MEGCF.
    \item NGCF and MMGCN show relatively flat performance on the three datasets \textit{w.r.t.} the number of layers, while LightGCN and MEGCF demonstrate a significant upward trend as the number of layers increases. We attribute such results to the fact that the nonlinear GCNs used by NGCF and MMGCN limit the capture of high-order CF signal, while LightGCN and MEGCF use linear GCN modules, which results in better performance. These results demonstrate the outperformance of the linear GCN structure for capturing high-order CF signals.
    \item MMGCN generally outperforms NGCF, while NDCG@10 of MMGCN is significantly weaker than that of NGCF on the Art dataset. This is attributed to the fact that the additional item multimodal deep features incorporated in MMGCN can enhance the user preference modeling. However, these features contain a considerable amount of preference-independent multimodal information, especially in the Art dataset where the negative impact of this information is greater. These results illustrate that the user preference-independent multimodal features can seriously contaminate embedding generation and ranking preference modeling.
    \item MEGCF consistently outperforms LightGCN significantly, which is due to the fact that MEGCF not only captures high-order CF signals but also models high-order multimodal semantic correlations. This demonstrates the importance of high-order multimodal semantic correlation in modeling user preferences. Moreover, compared with MMGCN, MEGCF, which is also based on multimodal information, exhibits a more pronounced upward trend as the number of graph convolution layers increases. This demonstrates that modeling high-order multimodal semantic correlations is more effective than existing multimodal feature processing approaches.
\end{itemize}

\subsection{Case Study (RQ3)}\label{sbusubsec:ex:cs}
\begin{figure}
	\centering
	\includegraphics[height=7.03cm, width=10cm]{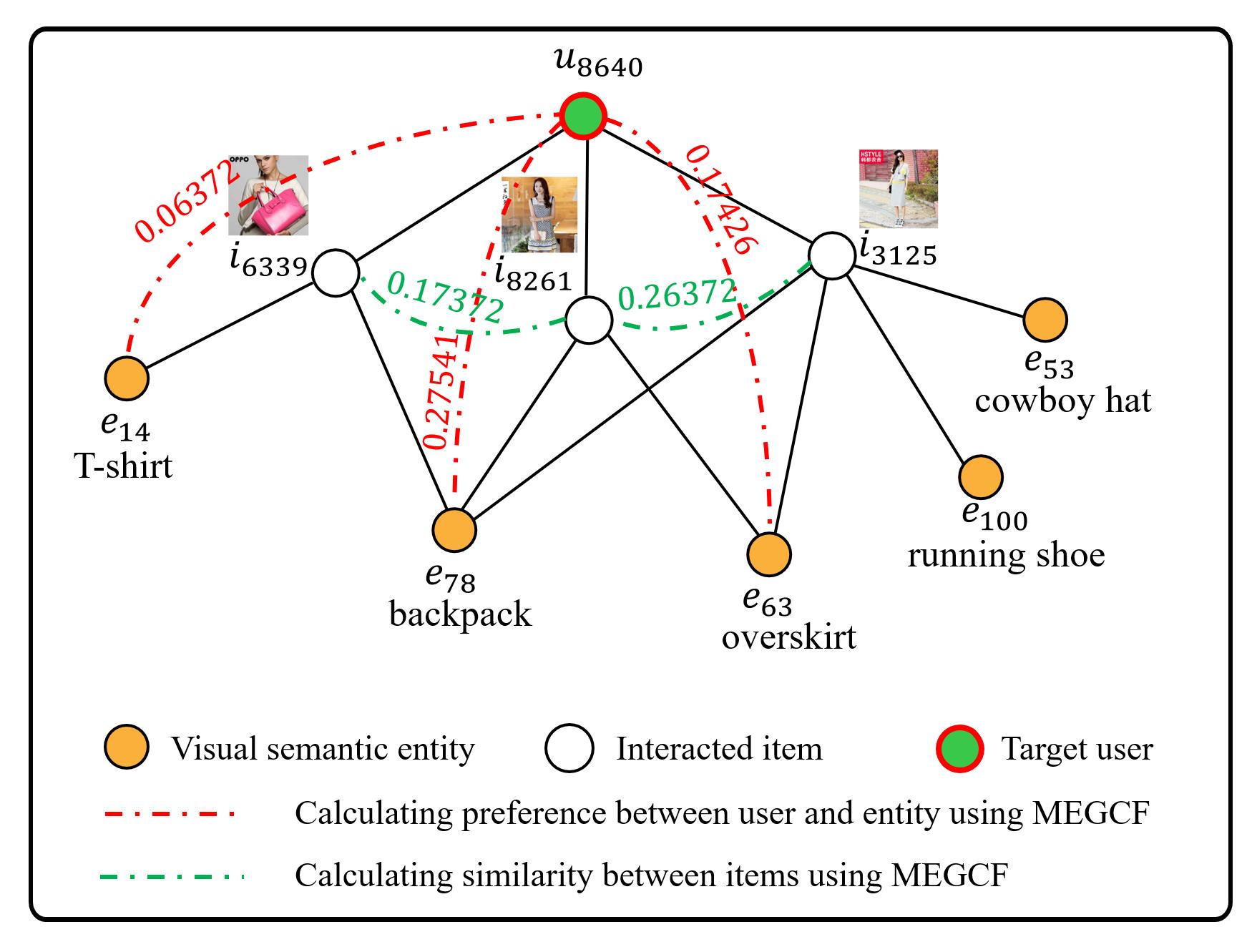}
	\caption{Real example from Taobao dataset, the target user is $u_{8640}$.}
	\label{fig:cs}
\end{figure}
In order to more intuitively explain the important role of multimodal semantic correlation in modeling modality-level user preference and item similarity, we design a simple case study on the Taobao dataset as shown in Figure \ref{fig:cs}. For simplicity, we only consider semantic entity mining of visual modality. Firstly, we randomly select a user $u_{8640}$ having an ID of 8640. Her historical interacted items are $i_{6339}$, $i_{8261}$, and $i_{3125}$, and we give the images corresponding to these items in Figure \ref{fig:cs}. Then, we leverage the method introduced in Section \ref{subsec:mmem} to mine the visual semantic entities in the images.
Afterwards, using all this information, a partial collaborative multimodal interaction graph with user $u_{8640}$ as the central node can be constructed. Finally, we use the MEGCF that has been trained on the Taobao dataset to predict user preference scores for different modality semantic entities, and calculate the similarities between different items using the inner product of the embeddings. We have the following findings:
\begin{itemize}[leftmargin=*]
    \item For the semantic entities ($e_{14}$: T-shirt, $e_{78}$: backpack, and $e_{63}$: overskirt), MEGCF computes a significantly higher score for $e_{78}$ than for $e_{14}$ and $e_{63}$, which is consistent with the fact that the images of all the three items contain $e_{78}$, \textit{i.e.}, this user is more interested in the "backpack". This result illustrates that incorporating multimodal semantic entities can effectively model user preference over multimodal latent space.
    \item The similarity score between items $i_{8261}$ and $i_{3125}$ is higher than that between items $i_{6339}$ and $i_{3125}$, \textit{i.e.}, compared with $i_{6339}$, $i_{8261}$ is more similar to $i_{3125}$, which also corresponds to the fact that $i_{8261}$ and $i_{3125}$ share three semantic entities, while $i_{8261}$ and $i_{6339}$ share only two. This result illustrates that capturing multimodal semantic correlation can better mine the similarity between items at the multimodal level.
    \item Based on this simple case, we find that there is still room for enhancing the multimodal semantic entity extraction in MEGCF. Specifically, since the deep method we leverage for visual semantic entity mining is pre-trained on the ImageNet dataset with only 1000 categories, the extracted semantic entities are limited by these 1000 categories. For example, all the styles of hats and bags in the images are roughly identified as "cowboy hat" and "backpack", respectively. In addition, directly transferring the deep approach from the computer vision research field to the recommendation scenario would significantly reduces the accuracy of the semantic entity recognition. Nevertheless, MEGCF still achieves the state-of-the-art performance. Based on this analysis, MEGCF will be significantly enhanced if the categories of entities can be refined and the accuracy of multimodal semantic entity extraction can be improved.
\end{itemize}

\section{CONCLUSION AND FUTURE WORK}
In this work, we propose a novel GCN-based multimodal recommender method, referred to as MEGCF, which introduces multimodal semantic correlation to tackle the mismatch problem in multimodal recommendation scenarios. By constructing a symmetric linear graph convolution network, MEGCF can achieve simultaneous capture of high-order multimodal semantic correlation and collaborative signal. In addition, we design a review-based sentiment weighting strategy to enhance the neighbor aggregation in GCN-based methods, in order to better capture high-order structural features on the graph. We conduct extensive experiments on three real-world datasets. The obtained results demonstrate the state-of-the-art performance of MEGCF. Further ablation experiments and analysis validate the effectiveness and rationality of MEGCF.

The multimodal entity extraction and semantic correlation modeling in MEGCF still has room for improvement as follows:
\begin{itemize}[leftmargin=*]
    \item Improving the accuracy of the extraction of multimodal semantic entities can better model semantic correlation and thus reduce the negative impact of misidentified entities. Consequently, in future work, we aim at using a larger dataset to enhance the pre-training of the feature extraction module, while employing contrastive learning \cite{sgl-sigir2021} techniques to improve the representation learning of multimodal features in a self-supervised manner.
    \item Multimodal semantic entity extraction in MEGCF fails to extract complete preference-related semantic information, thus leading to the limitation of low utilization of multimodal features. To tackle this problem, we aim at using techniques such as causal inference\cite{clickbait-sigir2021} and transformer\cite{attention1} in order to discover and distinguish more fine-grained preference-related multimodal information in future research.
    \item Rich semantic information in multimodal content can also be used to enhance the interpretability of recommender systems, which is left for future work.
\end{itemize}

\begin{acks}
	This work is supported by the Seventh Special Support Plan for Innovation and Entrepreneurship in Anhui Province, the Anhui Provincial Major Science and Technology Project under Grant 202203a05020025, and the National Natural Science Foundation of China under Grant 61876058.
\end{acks}

\bibliographystyle{ACM-Reference-Format}
\bibliography{sample-manuscipt}

\end{document}